\documentclass[twocolumn,showpacs,preprintnumbers,amsmath,amssymb]{revtex4-1}
\usepackage[utf8]{inputenc}
\usepackage{siunitx}
\usepackage{romannum}
\usepackage{hyperref}
\usepackage[capitalise]{cleveref}

 \newcommand{\me}{\mathrm{e}}
\newcommand{\mi}{\mathrm{i}}
\renewcommand{\Im}{\text{Im}}
\renewcommand{\Re}{\text{Re}}
\usepackage{todonotes}

\begin{document}

\title{Coulomb interactions and effective quantum inertia of charge carriers in a macroscopic
conductor}

\author{A. Delgard$^{1}$, B. Chenaud$^{1}$, 
U. Gennser$^{2}$, A. Cavanna$^{2}$, D. Mailly$^{2}$, P. Degiovanni$^{3}$ and C. Chaubet$^{1}$}
\affiliation{$^{1}$Universit\'e Montpellier 2, Laboratoire Charles Coulomb UMR5221, F-34095, Montpellier, France\\
CNRS, Laboratoire Charles Coulomb UMR5221, F-34095, Montpellier, France}
\affiliation{$^{2}$ Centre de Nanosciences et de Nanotechnologies (C2N), CNRS, 
Universit\'e Paris-Sud, Universit\'e Paris-Saclay, 91120 Palaiseau, France} 
\affiliation{$^{3}$ Univ Lyon, Ens de Lyon, Universit\'e
Claude Bernard Lyon 1, CNRS,
Laboratoire de Physique, F-69342
Lyon, France.}

\email[Corresponding Author : ]{christophe.chaubet@umontpellier.fr}

\begin{abstract}
We study the low frequency admittance of a 
quantum Hall bar of size much larger
than the electronic coherence length. We find that this macroscopic conductor 
behaves as an ideal quantum conductor with vanishing longitudinal
resistance and purely inductive behavior up to $f\lesssim \SI{1}{\mega\hertz}$.
Using several measurement configurations, we study the dependence of
this inductance on the length of the edge channel and on the integer
quantum Hall filling factor. The experimental data are
well described by a scattering model for
edge magnetoplasmons taking into account effective long range Coulomb interactions
within the sample. We find that the inductance's dependence on the
filling factor
arises predominantly from the effective quantum inertia of charge carriers
induced by Coulomb interactions.
\end{abstract}


\maketitle

By demonstrating that 
macroscopic conductors
could exhibit robust d.c. transport properties of quantum origin, the 
integer quantum Hall effect
(IQHE)~\cite{vonKlitzing-1980-1,Halperin-1982-1,Haug-1993-1,Weiss-2011-1,Suddards-2012-1} has been a major
surprise. The importance of this breakthrough 
for metrology was acknowledged
immediately~\cite{vonKlitzing-1980-1} and 
has lead to the redefinition of the 
Ohm~\cite{Ohm-definition-1988}. 
The finite frequency response
of quantum Hall conductors has been intensively studied
by
metrologists:
the use of an a.c. bridge at finite frequency $f$ revealed
departure of the Hall resistance $R_H(f)$ at $\nu=2$ from the expected value
$R_K/2=h/2e^2$~\cite{Jeckelmann-2001-1,Ahlers-2009-1,Delahaye-1995-1,Chua-1999-1,Schurr-2005-1}. 
It was then
attributed to “intrinsic inductances and capacitances”
\cite{Cage-1996-1,Jeanneret-1995-1}. Later, Schurr {\it et al} proposed a double
shielded sample allowing for a frequency-independent resistance
standard~\cite{Schurr-2011-1}, but these works left open the
question of the origin of these capacitances and inductances.

On the other hand, the finite frequency transport properties of quantum
coherent conductors, of size smaller than the electron coherence length,
are expected to be dominated by quantum effects. 
For low-dimensional conductors such as carbon nanotubes
\cite{Burke-2002-2}, or graphene \cite{Kang-2018-1},
the inductance is of purely kinetic origin. 
Small superconducting inductors
\cite{Annunziata-2010-1,Luomahaara-2014-1} now used in 
space industry \cite{Coiffard-2016-1} are based on the inertia of Cooper pairs. 
For a quantum coherent conductor, 
the theory developed by
B\H{u}ttiker and his collaborators
\cite{Buttiker-1993-1,Buttiker-1993-3,Pretre-1996-1} 
relates the associated $L/R$ or $RC$
times to the Wigner-Smith time delay for charge carriers scattering across the conductor.
These remarkable predictions have been confirmed by the measurement of
the finite frequency admittance of quantum Hall
R-C~\cite{Gabelli-2006-1} and R-L~\cite{Gabelli-2007-1,Song-2018-1} circuits 
of $\si{\micro\meter}$-size in the
$\si{\GHz}$ range at cryogenic temperatures. 

In this letter, we demonstrate that, in the a.c. regime, a 
$\si{\milli\meter}$ long ungated macroscopic quantum Hall bar, of
size much larger than the electronic coherence length,
exhibits a finite
inductance as well as a vanishing
longitudinal resistance. Such a purely inductive longitudinal response is expected
for quantum
conductors with zero backscattering: a kinetic energy cost
proportional to the square of the current 
arises from both the Pauli principle and the linear
dispersion relation for electrons close to the Fermi level (see
Appendix \ref{appendix/inductance}).
This
effective inertia of 
carriers causes the current response to lag the applied electric
field. Here, we
identify an inductance of the order of tens of
$\si{\micro\henry\per\milli\meter}$ and
connect it to an effective velocity $v_{\text{eff}}$ along the quantum
Hall bar's edges. Contrary to gated samples, in which $v_{\text{eff}}$ is
almost independent of the filling factor $\nu$ \cite{Gabelli-2007-1}, we show that,
because of Coulomb interactions between opposite edges of the sample, 
$v_{\text{eff}}$ depends on $\nu$ in our samples. Using the edge-magnetoplasmon
scattering approach combined with a discrete element approach {\it à la}
B\H{u}ttiker, we show that:
\begin{equation}
\label{eq:v0}
	\frac{v_{\text{eff}}(\nu)}{v_{\text{d}}(\nu)}=1+\frac{\nu\alpha_{\mathrm{eff}}(\nu)}{\pi}
	\ln\left[\frac{W/\xi_H(\nu)}{\nu}\right]\,
\end{equation}
for a sample of width $W$.
Here, $v_{\text{d}}(\nu)$ represents the charge density wave velocity
along the system of $\nu$-copropagating chiral edge channels, neglecting
Coulomb interactions with the other (counter-propagating) edge channels.
In a
Büttiker view of the edge channels \cite{Buttiker-1988-1},  $v_d(\nu)$ is
the drift velocity of non-interacting electrons at the edge in an
effective confining potential $U_\nu$ and, therefore, it
plays the role of
an effective Fermi velocity in the 1D linear dispersion relation along the edge
\cite{Book-Girvin}. In the presence of compressible
stripes, which appear for a sufficiently smooth confining potential
\cite{Chklovskii-1992-1},
it corresponds to the effective velocity of the charge density mode
in the system of $\nu$ copropagating edge channels
\cite{Aleiner-1994-1},
taking into account the presence of the incompressible part of the edge
channel
\cite{Han-1997-2}. Nevertheless, we denote it by $v_d(\nu)$ because,
in a model of
an edge channel without compressible parts, it really would be an electronic drift
velocity. Importantly,
this velocity deviates from
from the classical $1/B$ behavior of the electronic drift velocity
because of screening effects, which change the electrostatic
potential at the edge as $\nu$ varies.

Here, $\alpha_{\mathrm{eff}}(\nu)$
denotes the effective
fine structure constant ($\alpha_{\text{qed}}$ in the vacuum)
at filling factor $\nu$:
$\alpha_{\mathrm{eff}}(\nu)=
(\alpha_{\text{qed}}/\varepsilon_r)\times (c/v_{\text{d}}(\nu))$. The length 
$\xi_H(\nu)$, which also depends on $\nu$, 
is an effective renormalized width of a single edge channel
of the order of 
the width of incompressible edge channels $\lambda_H(\nu)$ %=\hbar/m^*v_{\text{d}}(\nu)$
\cite{Chklovskii-1992-1} (see Appendix \ref{appendix/capacitance})

Our work demonstrates that the purely inductive response of
the macroscopic ungated quantum Hall bar reflects the effective quantum inertia
of charge carriers renormalized by Coulomb interactions within the sample. Therefore, although
electron transport across such a conductor is not coherent, its d.c. and
a.c. transport properties are of quantum origin, a fact that ultimately
relies on the coherence of edge-magnetoplasmon (EMP) modes propagating along
chiral edge channels. EMP coherence has enabled the demonstration of single and double
EMP Fabry-P\'{e}rot interferometers \cite{Hashisaka-2013-1} as well as
of a Mach-Zehnder plasmonic
interferometer \cite{Hiyama-2015-1}.

Using shallow etching, our samples are processed on an AlGaAs/GaAs heterojunction with
the two dimensional electron gas (2DEG) located at the hetero-interface
($\SI{105}{\nano\meter}$ beneath the surface)
with carrier density $n_s=\SI{5.1e11}{\centi\meter^{-2}}$
and mobility $\mu=\SI{30}{\meter^2\per\volt\second}$.
We have processed a $2\times 0.4\si{\milli\meter^2}$
ungated Hall bar 
which exhibits a sufficiently large kinetic inductance. 
The sample has no back gate and is glued on a ceramic sample holder to avoid parasitic 
capacitances. It
is placed at the center of a high magnetic field 
at 
$\SI{1.5}{\kelvin}$. 
\begin{figure}
\includegraphics[width=0.9\columnwidth]{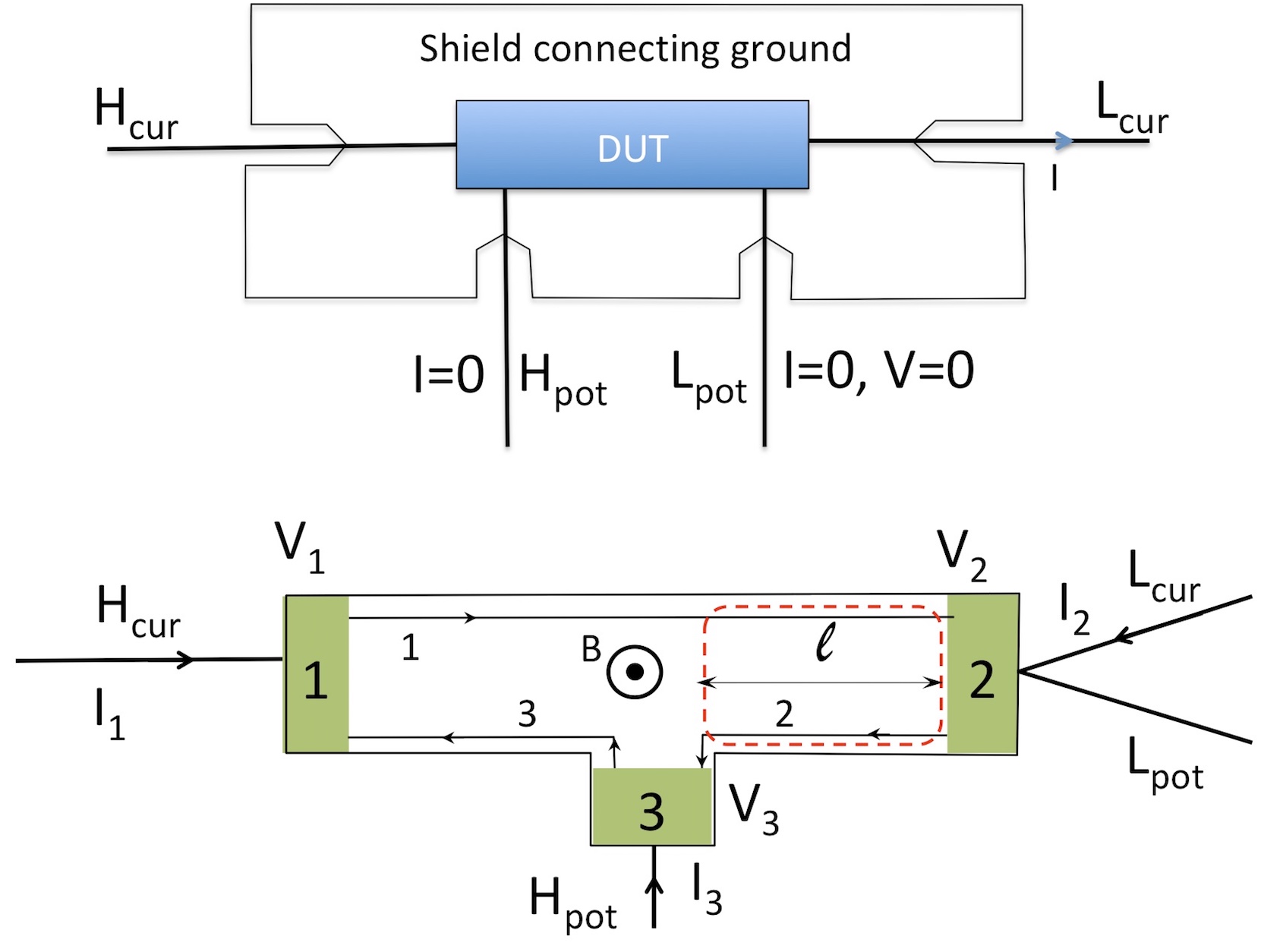}
\caption{ a) The Device Under Test (DUT) is measured using four coaxial
	cables and the impedance-meter Agilent HP4294A, which measures the
	current $I$ at $L_{\text{cur}}$, the potential
	$V$ at $H_{\text{pot}}$, and gives $G=I/V$,
	for details see Ref.~\cite{Book-Kibble-Rayner,Handbook-Agilent}.
	Note that the potentials of $H_{\text{pot}}$, $L_{\text{pot}}$,
	$H_{\text{cur}}$ and $L_{\text{cur}}$ are the potentials of the four
	connectors of the impedance-meter.
	b) Scheme of the
	multi-terminal Hall bar with only three ohmic contacts wire-bonded
	onto the sample holder. In this geometry, the impedance is
	$Z_{23}^{(\text{exp})}(\omega)=-(\partial
	V_3/\partial I_2)(\omega)$ for $V_2=0$.} 
	\label{figure1}
\end{figure}

In the measurement setup depicted on Fig.
\ref{figure1}-a., 
the current is injected using $H_{\text{cur}}$ ($\SI{5}{\milli\volt}$ bias),
and measured using $L_{\text{cur}}$. The potential of $H_{\text{pot}}$ is measured
while $V=0$ and $I=0$ are imposed at $L_{\text{pot}}$. 
The
current intensity ($\lesssim\SI{0.5}{\micro\ampere}$ at $\nu=2$) remains
below the breakdown current and 
currents used in
metrology \cite{Weiss-2011-1,Jeckelmann-2001-1}. For each values of $B$,
the resistance and the reactance have been measured for 
300 values of the frequency $f$ in the range
$\SI{40}{\hertz}$-$\SI{100}{\kilo\hertz}$.

Due to chirality of the quantum Hall transport, an ohmic contact
wire-bonded to the sample holder and so to a coaxial cable, generates a
leakage current through the cable capacitance if the potential does not
vanish~\cite{Grayson-2005-1, Hernandez-2014-1}. This results in a faulty
measurement~\cite{Desrat-2000-1,Melcher-2001-1}. For this reason, all results presented
here have been carried out at integer filling factors, where the
longitudinal resistance $R_{xx}(\omega)$ vanishes\footnote{Here $R_{xx}(\omega)$ denotes the frequency dependant
longitudinal dependence of the quantum Hall bar.}.
Furthermore, only $3$ of the ohmic
contacts processed on the sample were wire-bonded onto the sample
holder as shown on Fig.~\ref{figure1}.b. To measure a zero
resistance state, the third contact is inserted along the edge connected
to the reference potential. 
In d.c., one would measure a potential
$V_{\mathrm{H}_{\mathrm{pot}}}=0$. In a.c.,
$V_{\mathrm{H}_{\mathrm{pot}}}\neq 0$ and we measure the frequency
dependent impedance
$Z_{23}^{(\text{exp})}(\omega)=-\left. \partial V_3/\partial
I_2\right|_{V_2=0}(\omega)$.
Different configurations and edge channel lengths can be obtained by
recabling the contacts and changing
the sample side (in this case, the magnetic
field orientation must be reversed). 
We have also
wire-bonded a fourth ohmic contact on the same side of the
sample to connect $L_{\text{pot}}$ to access another edge
channel length.  

\begin{figure}[h!]
\includegraphics[width=0.9\columnwidth]{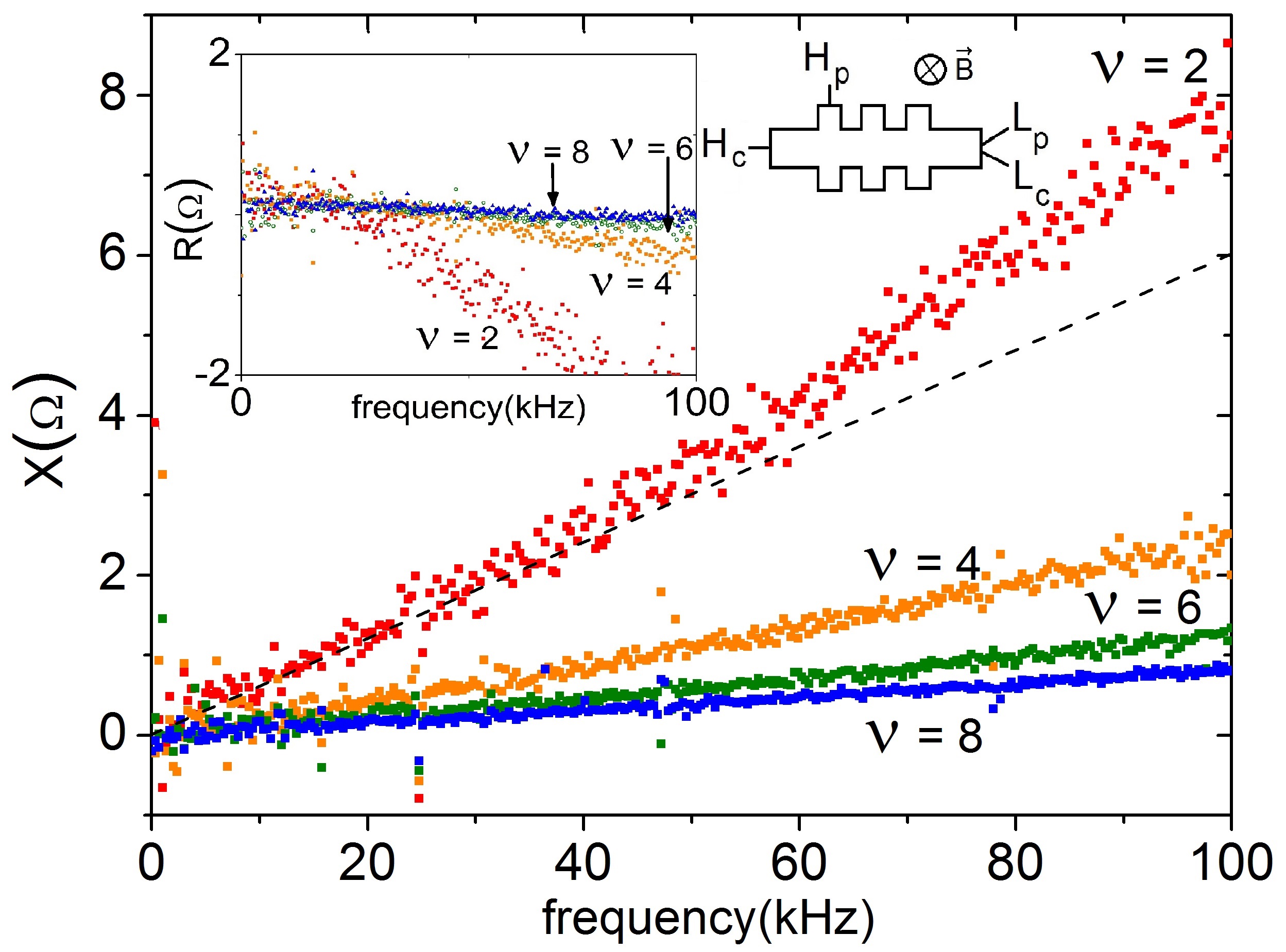}
	\caption{The reactance $X(f)$ as a function of the frequency $f$ for different
	$\nu$ and $B_{< 0}$, in the measurement configuration shown here.
	Inset: the longitudinal resistance $R(f)$ vanishes quadratically 
	for integer filling factors at low frequency. 
	}
	\label{figure2}
\end{figure}

Figure \ref{figure2} presents unfiltered and non-averaged raw data for
the reactance $X(f)=\Im(Z_{23}^{(\text{exp})}(2\pi
f))$ in a given sample
configuration for
$\nu=2$, $4$, $6$ and $8$.
The
positive linear dependence of $X(f)$ is the signature of an inductive
behavior. 
The corresponding inductance decreases with $\nu$.
These data are completely reproducible in the regions of magnetic fields
where $R_{xx}=0$.
This is a key point of our work: for integer filling factors, 
the real part $R(f)=\Re(Z_{23}^{(\text{exp})}(2\pi f))$ of the measured impedance
is close to zero with values
between $\pm \SI{0.5}{\ohm}$
at low frequency as shown in the inset of Fig.~\ref{figure2}. These results extend the
work of Gabelli {\it et al} \cite{Gabelli-2007-1} 
in which the sample resistance was $R_H=R_K/\nu$,
to the case of a zero resistance macroscopic device. At higher frequencies, a small 
real part of the measured response function $R(f)$ appears.
This effect
is discussed in Appendix \ref{appendix/emittances} and is correlated to the 
deviation of the reactance $X(f)$ from linearity seen on Fig. \ref{figure2}. 

Since the ac transport properties of a quantum Hall conductor is
directly realted to the scattering
of edge-magnetoplasmons \cite{Safi-1995-1,Safi-1999-1,Bocquillon-2013-2}, 
we have developed an analytical model (see Appendix
\ref{appendix/plasmons})
in the spirit of Ref. \cite{Hashisaka-2013-1} for 
scattering of EMP modes in
a quantum Hall bar taking into account long range inter-channel Coulomb
interactions. 
It assumes that,
in our ungated quantum Hall bars, 
all edge channels of same chirality have the same velocity
$v_{\text{d}}(\nu)$ and are so strongly coupled that
they see the same time dependent potential as in Ref.
\cite{Song-2018-1}. Since edge states are distant from more than \SI{1.5}{\centi\meter} 
from shield of coaxial cables located beneath the sample holder, 
estimated parasitic capacitance to shield
for edge states is below $\SI{1}{\femto\farad}$ while the inter-edge
capacitance $c_H$ is of the order of 
$\SI{0.1}{\pico\farad}$. Therefore,
Coulomb interactions effects are dominated by the inter-edge
capacitance $C_H$.
Finally, dissipation of EMP modes have been neglected, an hypothesis a
posteriori satisfied in our samples.

In the geometry depicted on Fig. \ref{figure1},
the low frequency expansion of the measured
reactance is of the form:
\begin{equation}
	\label{eq:Z-low-frequency}
	\Im\left(Z_{23}^{(\text{exp})}(\omega)\right) 
	=\mi L\omega +\mathcal{O}(\omega^2) 
\end{equation}
where
$L$ denotes the total
inductance for the quantum Hall bar delimited by a dashed red box on
Fig. \ref{figure1}-b. 
Because here, the magnetic inductance
is much smaller than the kinetic inductance 
(see Appendix \ref{appendix/inductance}),
$L$ 
can be obtained from the edge
magnetoplasmon scattering model as
\begin{subequations}
	\label{eq:L}
	\begin{align}
	\label{eq:expression-basique-L}
		L= (R_K/\nu)\times(l /2\,v_{\text{eff}}(\nu)) \\
	\label{eq:expression-velocity}
		v_{\text{eff}}(\nu)=v_{\text{d}}(\nu)\times \left( 1+
		\frac{C_q(\nu)}{2C_H(\nu)} \right) \,.
		\end{align}
\end{subequations}
where $l$ is the length of the Hall bar (see Fig. \ref{figure1}-b),
$C_q(\nu)=\nu e^2l/ hv_{\text{d}}(\nu)$ is the quantum capacitance of $\nu$ edge channels
and the geometric capacitance $C_H(\nu)$
describes the effect of Coulomb interactions
between counter-propagating edge channels. This is different
from
the quantum RL-circuit where, because of the gating, 
the capacitance $C_H$ has to be replaced by the capacitance $C_g$ with
the nearby gates leading to a renormalization of $v_{\text{d}}$ by $1+C_q/C_g$
for right and left moving charge density waves \cite{Gabelli-2007-1}. Here, 
the
renormalization factor involves a $C_H$ capacitance with the
series addition $C_q/2$ of the quantum capacitances of counter-propagating edges.
As a consequence,
Eq. \eqref{eq:L} still relates $L=R_H^2C_\mu$ to the Hall
resistance and to the electrochemical capacitance \cite{Christen-1996-1} $C_\mu$ defined
as the series addition of $C_H$ to $C_q/2$.
Eq. \eqref{eq:L} suggests that the inductance can be
interpreted as a kinetic inductance associated with an effective time of flight 
$l/v_{\text{eff}}(\nu)$. 
But, as discussed in Appendix \ref{appendix/plasmons},
$v_{\text{eff}}(\nu)$ is neither the drift
velocity for non-interacting electrons nor
even a renormalized electron's velocity within chiral edge channels, but an
effective velocity 
arising from the combination of their kinetic
quantum inertia and Coulomb interactions within the quantum Hall bar.
This effective inertia is of quantum origin, reflecting the minimal energy
associated with an electrical current
and appears, as we will see, to be dominated 
by the effects of Coulomb interactions.

The geometric capacitance $C_H(\nu)$ depends on the 
width $W$ of the sample, and of the structure and geometry 
of the quantum Hall edge channels (see Appendix
\ref{appendix/capacitance}),
through 
 a length $\xi_H(\nu)$ proportional to the width $W_H(\nu)$ of a single channel. 
Following Ref.~\cite{Chklovskii-1992-1},
$W_H(\nu)=(1+\pi^2\alpha_{\text{eff}}(\nu))\lambda_H(\nu)$,
which is of the order of a few 
tens of $\si{\nano\meter}$ for AlGaAs/GaAs quantum
Hall systems. 
Finally, the inter-edge Coulomb interactions contribution to
$v_{\text{eff}}(\nu)$ 
\begin{equation}
	\label{eq/v0/subdominant}
	\frac{v_{\text{d}}(\nu)C_q(\nu)}{2C_H(\nu)}\simeq
	\frac{\sigma_H(\nu)}{2\pi\varepsilon_0\varepsilon_r}\,\ln\left(\frac{W}{\nu\xi_H(\nu)}\right)\,
\end{equation}
is found to be linear in $\nu$, because of its proportionality to
the quantum Hall conductivity $\sigma_H(\nu)$, but with a
logarithmic multiplicative correction which is a signature of Coulomb interactions. 

We will now discuss how
this expression
and the experimental data enable us to rule
out some models for $v_{\text{d}}(\nu)$.
We have considered two different models for the confining potential
at the edge of the sample which leads to a different prediction for
$v_{\text{d}}(\nu)$: in Ref.~\cite{Mikhailov-2000-1},
$v_{\text{d}(\nu)}=\omega_c/k_F$ where $\omega_c=eB/m^*$ is the cyclotron frequency
and $k_F=\sqrt{2\pi /n_S}$ the Fermi wave-vector. This leads to a
dependance $v_{\text{d}}(\nu)=v_d/ \nu$ whereas in Ref.~\cite{Chaubet-1998-1}, the
gradient of the potential is proportional to $\hbar \omega_c/l_B$ where
$l_B=\sqrt{\hbar/eB}$ is the magnetic length thereby leading to a scaling
$v_{\text{d}}(\nu)=v_d / \sqrt{\nu}$. 

\begin{figure}[h!]
\includegraphics[width=0.90\columnwidth]{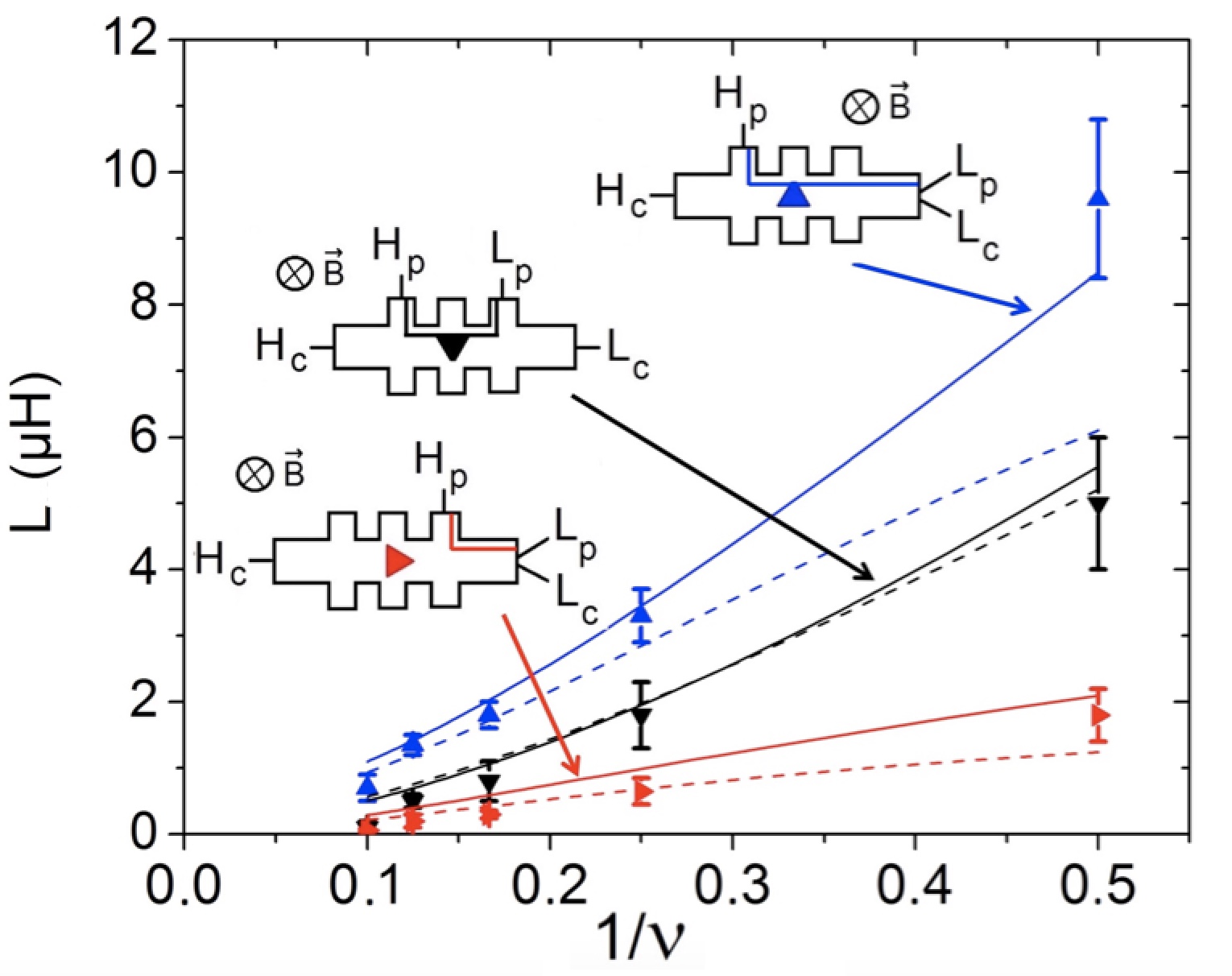}
	\caption{(Color online) For measurement configurations with
	$B<0$
	corresponding to 
	$l=\SI{600}{\micro\meter}$, 
	$\SI{1000}{\micro\meter}$ and $\SI{1600}{\micro\meter}$,
	the inductance increases
	with $1/\nu$.
	Dashed lines correspond to model $v_{\text{d}}(\nu)=v_{\text{d}}/\nu$ 
	with 
	 $v_{\text{d}}=15$, $5$ and $40$ (in units of
	 $\SI{e5}{\meter\second^{-1}}$) from top to bottom.
	The $v_{\text{d}}(\nu)=v_{\text{d}}/\sqrt{\nu}$
	model leads to (solid lines) $v_{\text{d}}=6$, $3$ and $17$ 
	in units of $\SI{e5}{\meter\per\second}$ from top to bottom.
	The blue
	points corresponds to the experimental data displayed on Fig.
	\ref{figure2}.
} 
\label{figure3}
\end{figure}

Fig. \ref{figure3} contains the first main quantitative result of this work, {\it i.e.} 
the quantum inductance as function of $1/\nu$ for
configurations $B<0$ ($B>0$ configurations are analyzed in 
Appendix \ref{appendix/extra-results}).
Values have been obtained from the reactance data 
depicted on Fig. \ref{figure2}
using the
slope at low frequency of $f\mapsto X(f)$ datasets.
Three configurations in which $L_{\text{pot}}$
and $H_{\text{pot}}$ are plugged to different contacts (see 
Fig.~\ref{figure3}) and thus correspond to different $l$
have been studied.
The main result is 
the dependence of the inductance on $1/\nu$ which involves a linear part
(see Eq. \eqref{eq:expression-basique-L}) due
to the presence of $\nu$ channels in parallel, but with a 
non-linear correction stemming from the $\nu$-dependence of $v_{\text{d}}(\nu)$
(see Eqs.~\eqref{eq:expression-velocity}) together with Coulomb
interaction effects (see Eq.~\eqref{eq/v0/subdominant}). The different
dependences of $v_{\text{d}}(\nu)$ 
lead to different theoretical predictions. We find
that the scaling $v_{\text{d}}(\nu)=v_{\text{d}}/\sqrt{\nu}$ is the best for
describing the experimental data.

\begin{figure}[h!]
\includegraphics[width=0.90\columnwidth]{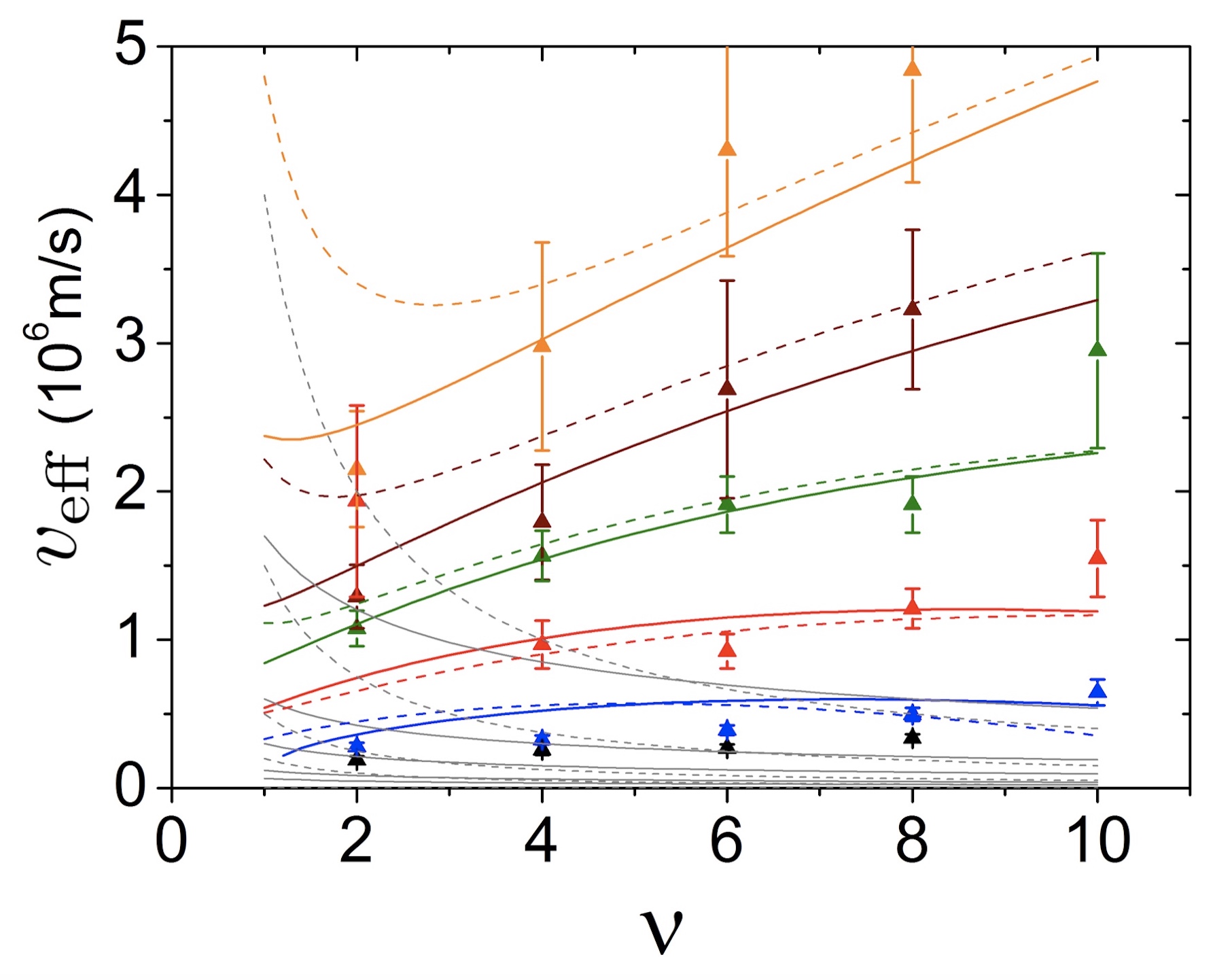}
\caption{(Color online) 
	$v_{\text{eff}}$ 
as a function of $\nu$. Dashed curves correspond to
$v_{\text{d}}(\nu)=v_{\text{d}}/\nu$ and full lines to $v_{\text{d}}(\nu)=v_{\text{d}}/\sqrt{\nu}$.
Colored lines correspond to the
expressions \eqref{eq:v0} taking into account inter-edge Coulomb
	interactions.
Thin grey curves correspond to plots of the bare velocity
$v_{\text{d}}(\nu)$. The full lines have been obtained with (starting from
the top curve):
$v_{\text{d}}= 17$, 
	$v_{\text{d}}=6$ (fit of the data from Fig.
	\ref{figure2}),
$v_{\text{d}}= 3$,
$v_{\text{d}}= 1.2$ and
	$v_{\text{d}}= 0.66$ in units of $\SI{e5}{\meter\per\second}$. For the dashed lines, we have
starting from the top curve:
$v_{\text{d}}=40$,
$v_{\text{d}}=15$,
$v_{\text{d}}=5$,
$v_{\text{d}}=2$ and
$v_{\text{d}}=1$ in units of $\SI{e5}{\meter\per\second}$.
	}
	\label{figure4}
\end{figure}

We have then
extracted the velocity $v_{\text{eff}}(\nu)$ using Eq.
\eqref{eq:expression-basique-L} from each value of the inductance (see
Fig. \ref{figure4}).
This is the second main quantitative experimental result of this work. 
Each family
of points corresponds to a specific sample configuration for which the sample
has been heated up, re-bonded and cooled down again. These manipulations
affect the electrostatic arrangement of charges at the edge,
thereby 
modifying $v_{\text{d}}$ from one experiment to the other. 
Fig. \ref{figure4}
presents predictions for $v_{\text{eff}}(\nu)$ from Eq.~\eqref{eq:v0}
for different models for $v_{\text{d}}(\nu)$.
Simarily to the discussion of Fig.
\ref{figure3}, the $\nu^{-1/2}$ scaling for $v_{\text{d}}(\nu)$ gives the best 
reproduction of the experimental data. But a striking point is that
in order to reproduce
the experimental data, it is necessary to 
take into account the inter-channel
Coulomb interactions: ignoring the interactions would correspond to using $v_{\text{d}}(\nu)$ instead
of $v_{\text{eff}}$ in Eq. \eqref{eq:expression-basique-L}. This is
shown by the thin filled and dashed
grey curves on Fig. \ref{figure4}, which clearly do not
follow the experimental data. We thus interpret the
$\nu$-dependence of 
$v_{\text{eff}}(\nu)$ when increasing 
$\nu$ from $2$ to $10$ (mostly linear with $\log$-correction) as a
strong indication of the dominant role of Coulomb interactions in these ungated samples.

Let us comment on the spread of values for $v_{\text{d}}$ given
in Fig.~\ref{figure4}, which
reflects the variability of the electrostatic environment in the
samples from one experimental cooldown to another.
A variation by a
factor 25 is observed across all
experiments (three higher curves for $B>0$, all others for $B<0$) but by
only a factor 6 when considering only one orientation of $B$. This is
still much
larger than the relative change of the 2DEG density but $v_{\text{d}}$
reflects the edge potential, which may vary more from one
experiment to the other. As we use a shallow etching technique, the
samples edges are very sensitive to any change of the electric potential
landscape\footnote{Additional results obtained on
samples from different wafers are presented in
Appendix \ref{appendix/extra-results}
provide additional evidence of
the robustness of our analysis.}. The values that we have obtained for
$v_{\text{d}}(\nu=2)$ are compatible with estimates in the literature
\cite{Roulleau-2008-2} for shallow etched samples. Moreover, our
measurements of
$v_{\text{eff}}(\nu)$ are in the same range and qualitatively show a
similar $\nu$-dependence as the ones obtained in Ref.
\cite{Kumada-2011-1} for ungated samples by a time-of-flight technique.

To summarize, we have shown that, at low frequencies, a macroscopic 
quantum Hall bar is a
perfect 1D conductor exhibiting a vanishing longitudinal resistance and a finite
inductance. By fitting its dependence on $\nu$ and on
the sample geometry using a simple long range effective Coulomb
interaction model in the spirit of Büttiker {\it et al}
\cite{Christen-1996-1}, 
we have shown that it reflects the effective
quantum inertia of charge carriers within the edges
of the quantum Hall bar.
Contrary to the case of superconductors where
carrier inertia arises from the effective mass of the Cooper pairs,
here it reflects how Coulomb
interactions alter the propagation of
low-frequency massless edge-magnetoplasmon modes. 
Remarkably, the experimental data
can be understood using a simple model which is
formally similar to the one used for gated nano-fabricated samples
\cite{Hashisaka-2012-1,Hashisaka-2018-1}: starting 
from chiral charge transport
with bare velocity $v_{\text{d}}(\nu)$, we include
Coulomb interactions with the other edges
via classical electrostatics and edge structure geometry from Ref.
\cite{Chklovskii-1992-1}.
Going beyond this phenomenological but practical model would involve
a multiscale treatment combining our approach to inter-channel
interactions with a self-consistent
microscopic approach 
solving the problem of electrons in the presence of
intra-channel Coulomb interactions
\cite{Armagnat-2019-1,Armagnat-2020-1,Roussely-2018-1}. 

Finally, macroscopic samples may provide a rescaled test-bed for studying the scattering
properties of edge-magnetoplasmons in $1$ to $\SI{100}{\micro\meter}$-sized samples.
Studying a.c. transport properties
of macroscopic samples up to radio-frequencies could thus open
the way to realizing controlled quantum linear components for quantum
nano-electronics in 1D edge channels, with possible 
applications to electron \cite{Bocquillon-2014-1} and 
micro-wave quantum optics in ballistic
quantum conductors \cite{Grimsmo-2016-1}. 

\begin{acknowledgements} 
We warmly thank G. Fève, Ch. Flindt and B. Pla\c{c}ais for useful discussions and suggestions 
as well as the referees for important
comments and clarifications on this manuscript.
This work has been partly supported by ANR grant ``1shot
reloaded'' (ANR-14-CE32-0017) and the French Renatech network.
\end{acknowledgements}

\appendix
\section{Kinetic and magnetic inductance}
\label{appendix/inductance}

In this section, we compare the kinetic inductance of a quantum Hall bar at filling
fraction $\nu$ to its geometric inductance. 

\subsection{Kinetic inductance}
\label{appendix/inductance/kinetic}

\begin{figure}[!h]
    \centering
    \includegraphics[width=8cm]{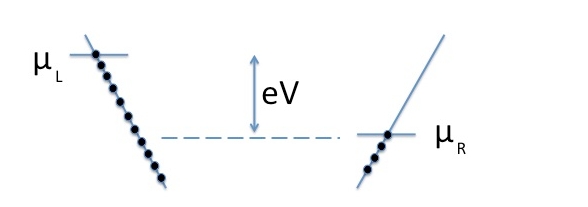}
\caption{\label{fig:supplementary:1} Dispersion relations for
non-interacting electrons propagating along a right-moving chiral edge
channel $R$ and a left moving one $L$ with propagation velocity $\pm
v_{\text{d}}$. In the presence of a d.c. voltage bias $V$, a chemical potential 
imbalance $-eV$ appears between the two
chiral edge channels. At filling fraction $\nu$, such 
a bias generates a net current $I=(e^2\nu/h)\times
V$.}
\end{figure}

Let us first consider non-interacting chiral charge 
carriers with propagation velocity $v_{\text{d}}$ (see Fig. \ref{fig:supplementary:1}). 
As explained in the main text, for incompressible quantum Hall edge channels, this
velocity is a drift velocity $v_{\text{d}}(\nu)=-\partial_yU_\nu/B$ %($U_\nu$ being
%the confining potential for the electrons), 
which depend on $\nu$ via the magnetic field and 
the dependence of the confining
potential $U_\nu$ on $\nu$ itself as will be detailed in Sec. \ref{appendix/nu-potential}. 
However, to keep notations simple, 
we shall denote it by $v_{\text{d}}$ instead of
$v_{\text{d}}(\nu)$ unless necessary.

A net current flow corresponds to a chemical potential between chiral
edge channels of opposite chirality.
The corresponding kinetic energy is then obtained by adding together the energy of all quantum states which have 
been occupied when biasing chiral edge states with different chemical
potential $\Delta \mu= -eV$:
\begin{equation}
	E_{\mathrm{chir}}=\sum_k \varepsilon(k)=\nu \times
	g(\varepsilon)l\times (e V )^2/2\,,
\label{Eq1}
\end{equation}
where $g(\varepsilon)= 1/ h v_{\text{d}}$ is the density of states (DOS) per unit length for a 1D-channel, $\nu$ 
the number of channels, $l$ the edge-state length. 
The voltage drop across the sample is $V=R_HI_\nu$, where
$R_H=R_K/\nu=h/e^2\nu$ denotes the quantum Hall resistance at filling
fraction $\nu$ and  
$I_\nu$ the current. The resulting kinetic energy is then 
quadratic in the current, as expected for a linear inductor:
\begin{equation}
\label{Eq2}
	E_{\mathrm{chir}}(I_\nu)= \frac{\nu e^2}{h v_{\text{d}}} \times \frac{l}{2}
	\times  \left(\frac{R_K I_\nu }{\nu}\right)^2 
= \frac{h}{e^2} \times \frac{l}{\nu v_{\text{d}}} \times \frac{I_\nu^2
	}{2}\,.
\end{equation}
For a quantum Hall bar, we have two counter-propagating groups of $\nu$ chiral
edge channels. Assuming that the two chiralities carry equal charge of
the total current $I_\nu=\pm I/2$, this leads to a total kinetic
energy
\begin{equation}
	E_K(I)=2E_{\text{chir}}(I/2)=\frac{h}{e^2}\times \frac{l}{\nu v_{\text{d}}}\times
	\frac{2(I/2)^2}{2}\,.
\end{equation}
Using $E_K(I)=L I^2/2$, we obtain the following formula for the
inductance $L$ of the quantum Hall bar:
\begin{equation}
	\label{eq:kinetic-inductance}
	L = \frac{R_K}{\nu} \times \frac{l}{2v_{\text{d}}(\nu)} 
\end{equation}
Note the appearance of the half of the time of flight $l/v_{\text{d}}(\nu)$ of the electrons
across the quantum Hall bar which is ultimately due to the existence of
two counter-propagating groups of copropagating edge channels.

\subsection{Magnetic inductance}
\label{appendix/inductance/magnetic}

A quantum Hall bar can be viewed as a conductor built from 
two linear wires of length $l$, and
of diameter or width $W_H$
separated by a distance $W$ which is the width of the quantum Hall bar. The magnetic
inductance of the resulting dipole can be computed from Biot-Savart's
law \cite{Rosa-1908-1}. Note that the wire's diameter $W_H(\nu)$ usually scales as
$\nu$
(see Ref.
\cite{Chklovskii-1992-1} and Sec. \ref{appendix/edge-channel-width}). 
In the quantum Hall bar
geometry, the total current $I$ flowing across the quantum Hall bar corresponds to
$\pm I/2$ along each of the two long counter-propagating edges.
Consequently, in the limit $l,W\gg W_H(\nu)$, the geometric inductance of the
quantum Hall bar is approximately equal to
\begin{equation}
	L_{\text{m}}=\frac{l\mu}{\pi} \left[\ln\left(\frac{2W}{W_H(\nu)}\right)-1 \right]
\end{equation}
where $\mu=\mu_r\mu_0$ is the magnetic permeability of the material
expressed as the product of its relative
permeability $\mu_r$  by
the vacuum permittivity $\mu_0=1/\varepsilon_0c^2$. For a
non-magnetic material such as AlGaAs/GaAs, $\mu_r=1$ and therefore
$\mu=\mu_0$.

Using this expression and Eq.~\eqref{eq:kinetic-inductance}, the ratio of the
magnetic to the kinetic inductance is given by:
\begin{equation}
	\frac{L_\text{m}}{L_K}=4\alpha_{\text{qed}}\,\frac{v_{\text{d}}}{c}\times
\nu
	\times\left[\ln\left(\frac{2W}{W_H(\nu)}\right)-1\right]
\end{equation}
where $\alpha_{\text{qed}}$ is the fine structure constant. The magnetic 
character of this ratio appears through
the multiplication factor $v_{\text{d}}/c\ll 1$. Together with 
the smallness of $\alpha_{\text{qed}}$ and the
logarithmic dependence in the aspect ratio $W/W_H(\nu)\sim W/\nu W_H(1)$, this
explains why the magnetic inductance is always much smaller than the
kinetic one for $\nu\lesssim 100$.

\section{Experimental signals} 
\label{appendix/emittances}

In this section, we discuss how the quantities that are measured in
the experiment are related to the a.c. response of a quantum Hall bar
even in the presence of experimental imperfections such as cable
capacitances.

\subsection{Finite frequency transport in quantum Hall edge channels}
\label{appendix/emittances/Christen-Buttiker}

In his pioneering work on finite frequency charge transport in
mesoscopic conductors \cite{Buttiker-1993-1}, Büttiker stressed the importance of
interactions: time dependent drives applied to reservoirs lead to charge
pumping in the conductor which, in return, alter the electrical
potential within the conductor. As a result, its transport properties
can no longer be derived by using the d.c. response at zero bias. Büttiker 
then developed a self consistent mean-field approach to this problem, taking into
account how the time dependent charge density within the conductor
generates a time-dependent potential and thereby alters the transport
\cite{Pretre-1996-1}.
In the linear regime, the a.c. response of the
conductor is described by the admittance matrix
\begin{equation}
\label{eq:def:admittance}
G_{\alpha\beta}(\omega)=\left.\frac{\partial I_\alpha(\omega)}{\partial
	V_\beta}\right|_{V=0}
\end{equation}
giving the average current entering the
conductor from lead $\alpha$ when a voltage drive at the same frequency
$\omega/2\pi$ is applied to the reservoir $\beta$.
Charge conservation and gauge invariance imply
that the finite frequency admittance matrix
satisfy the general sum rules 
\begin{subequations}
\begin{align}
\sum_\alpha G_{\alpha\beta}(\omega)&=0\\
\sum_\beta G_{\alpha\beta}(\omega)&=0
\end{align}
\end{subequations}
which is ensured by total screening of effective Coulomb interactions.

At low frequency, this admittance matrix is expanded in $\omega$, the
zero-th order term being
its d.c. conductance $G_{\alpha\beta}^{(\text{dc})}$. The first order
term is related to the emittances introduced by Büttiker {\it
et al.} :
\begin{equation}
\label{eq:definition:emittances}
	G_{\alpha\beta}(\omega)=
	G_{\alpha\beta}^{(\text{dc})}-\mi \omega
	E_{\alpha\beta}+\mathcal{O}(\omega^2)\,.
\end{equation}
For quantum Hall edge channels,
Christen and B\H{u}ttiker \cite{Christen-1996-1} fully exploited the chirality of charge
transport to 
compute  the dc conductances $G^{(\mathrm{dc})}_{\alpha\beta}$ as well as the
emittances $E_{\alpha\beta}$ using a discrete element description of the
circuit. 

Here we will show how to describe the finite frequency transport using 
the building blocks of a discrete element description of quantum Hall bars and the finite frequency
admittance properties obtained from the edge-magnetoplasmon (EMP)
scattering approach.  
%The latter
%has been first pioneered in the context of quantum wires 
%\cite{Safi-1995-1,Safi-1999-1} before being used for quantum Hall edge
%channels \cite{Degio-2010-1} where it is instrumental for computing
%single electron decoherence \cite{Ferraro-2014-2,Cabart-2018-1}. 
The details of how the admittances are obtained from the EMP
scattering approach are described in Sec. 
\ref{appendix/plasmons}.

Depending on the sign of the diagonal emittances, the quantum
conductor under consideration exhibits a capacitive
($E_{\alpha\alpha}>0$) or inductive ($E_{\alpha\alpha}<0$) behavior.
Basic examples of these behaviors, involving two
different two-contact devices, have been given by Christen and B\H{u}ttiker 
\cite{Christen-1996-1}. 
A two-contact Hall bar is predicted to act as an 
inductance while a ring shaped sample (the so-called Corbino geometry) 
acts as a capacitance. Hence both geometries exhibit opposite
diagonal emittances, which can be 
understood in terms of reaction of the circuit to an injected charge.
Indeed, as shown on Fig.~1 of \cite{Christen-1996-1}
a charge injected by contact 1
into a Hall bar is transmitted to contact 2, while the charge induced by
inter-edge Coulomb interactions created at 
contact 2 is transmitted to contact 1. 
On the contrary, in a Corbino sample, the injected charge returns to
the contact it comes from, as does the image charge.
It was experimentally shown by Delgard {\it et al.} \cite{Delgard-2019-1} 
that Corbino emittances will
exhibit this predicted capacitive behavior. Here, we will highlight the inductive behavior
of quantum Hall bars in a multi-contact geometry that ensures vanishing longitudinal resistance. 

We shall now
discuss how, for two different sample configurations, 
the measured response function relates to the finite frequency 
admittance of the quantum Hall bar depicted as a red dashed rectangle in the
two forthcoming figures, taking into account capacitive leaks at the
contacts.

\subsection{Three contact geometry}
\label{appendix/emittances/3-contacts}

\begin{figure}[!h]
\centering
\includegraphics[width=8cm]{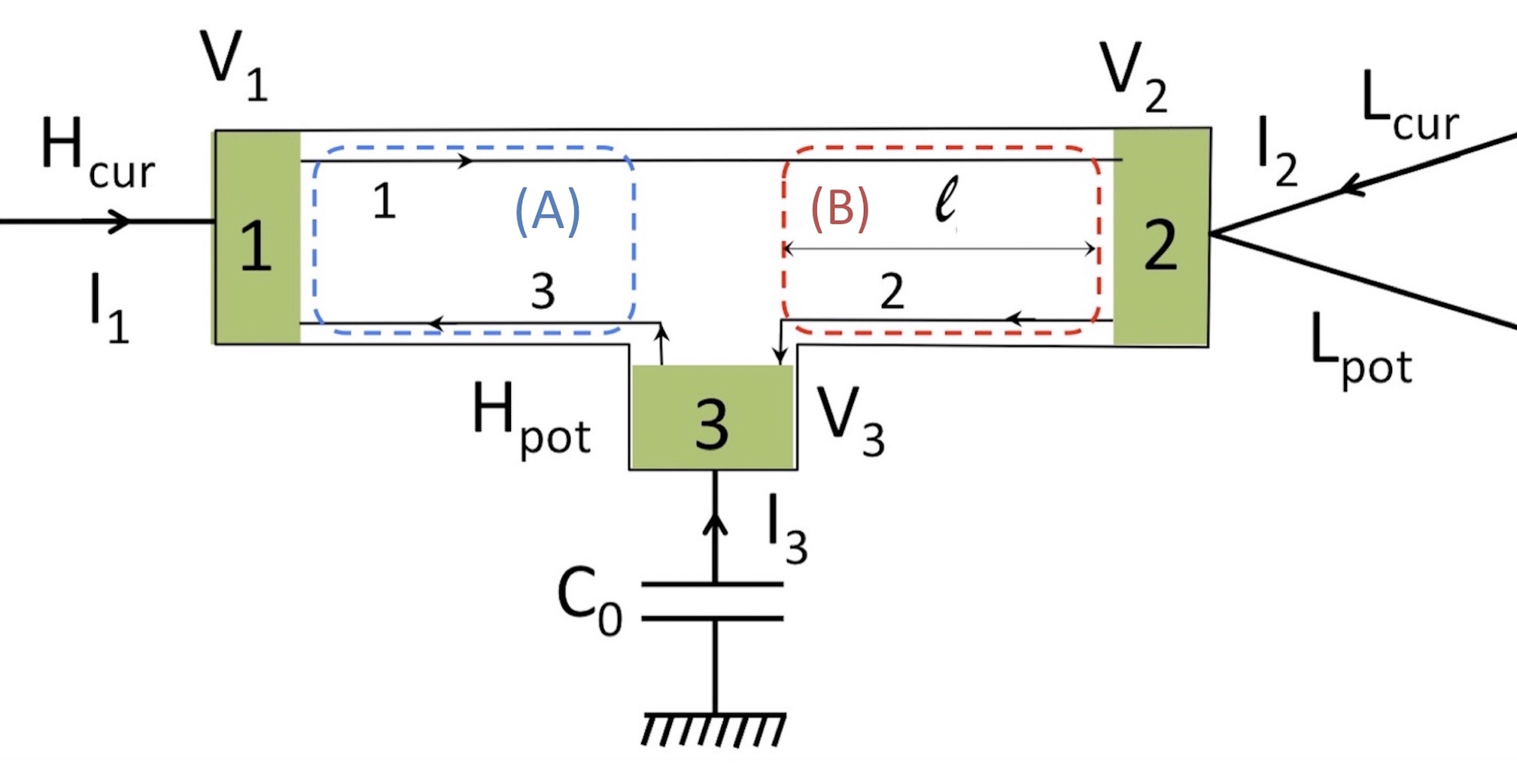}
\caption{\label{fig:3contacts} The device under test when
three Ohmic contacts are wire-bonded onto the sample holder. The
quantum Hall bar $B$, which is the dipole under test, 
is shown as a dashed red box. The quantum Hall bar $A$ is shown as a
dotted blue box. The capacitance of the coaxial cable connected to contact $3$ is denoted
by $C_0$. In this experimental setup, $V_2=0$ whereas the drive $V_1$ is
non zero and
the experimentally measured impedance is 
$Z^{(\text{exp})}_{23}=-\partial V_3/\partial I_2$.}
\end{figure}

Let us first consider the three contact geometry depicted on Fig. 
\ref{fig:3contacts}. Contacts $1$ and $2$ are directly connected to reservoirs whereas
contact $3$ is not but we take into account the cable capacitance $C_0$
of the cable.
In this section, we shall compute 
the experimentally measured finite frequency impedance 
\begin{equation}
 Z^{(\text{exp})}_{23}(\omega) =-\left.\frac{\partial V_3(\omega)}{\partial
		I_2(\omega)}\right|_{V_2=0}
\end{equation}
up to second order in $\omega$ and show how it relates to the
properties of the quantum Hall bar $B$ and
to the cable capacitance $C_0$ (see Fig. \ref{fig:3contacts}). 

\subsubsection{Obtaining $Z^{(\text{exp})}_{23}$}
\label{appendix/emittances/3-contacts/plasmon-scattering}

In this approach, the quantum Hall bars $A$ and $B$ delimited on Fig.
\ref{fig:3contacts} are characterized by 
their frequency-dependent EMP scattering matrix. For the quantum Hall bar $A$, the
EMP scattering matrix, which we denote here by $\mathbf{A}(\omega)$
relates 
the ingoing and outgoing currents through
\cite{Slobodeniuk-2013-1}
\begin{equation}
\mathbf{j}^{(\text{out})}(\omega) =
\mathbf{A}(\omega)\,\mathbf{j}^{(\text{in})}(\omega)
\end{equation}
in which the edges of the Hall bar are labeled by their chirality. 
In the same way, the EMP scattering matrix of the quantum Hall bar $B$
will be denoted by $\mathbf{B}(\omega)$.  
These matrices are computed in Sec.
\ref{appendix/plasmons} but for the moment, we won't need their explicit
form.

At contact $3$, the incoming and outgoing currents propagating within
the edge channels are also related by an input/output relation that
reflects the time delay associated with the $R_HC_0$ relaxation time of
the contact:
\begin{equation}
\label{eq-Ohmic-input-output}
i_{3,\text{out}}(\omega) =
\frac{i_{3,\text{in}}(\omega)}{1-\mi R_HC_0\omega}
\end{equation}
Eliminating the edge channel currents internal to the sample directly leads to
the finite frequency admittances connecting the currents $I_\alpha$
entering from the reservoirs to
the drive voltage $V_1$:
\begin{equation}
                R_H\,I_2 =
                -\frac{A_{RR}B_{RR}}{1-A_{RL}\mathcal{T}B_{LR}}\,V_1
\end{equation}
in which the matrix elements of the Hall bar's current scattering
matrices $\mathbf{A}$ and $\mathbf{B}$ are
involved as well as
$\mathcal{T}(\omega)=1/(1-\mi R_HC_0\omega)$, the edge current transmission of
the Ohmic contact with cable capacitance $C_0$. For simplicity, 
the $\omega$ dependence of $V_1$, $I_2$ and of
the scattering matrix elements of the $A$ and $B$ quantum Hall bars as
well as of $\mathcal{T}$ have been omitted.
The potential of the contact
$3$ can also be computed as
\begin{equation}
-\mi \omega C_0R_H\,V_3 =
\frac{(\mathcal{T}-1)B_{LR}A_{RR}}{1-A_{RL}\mathcal{T}B_{LR}}\,V_1\,
\end{equation}
which finally leads to
\begin{equation}
        \frac{Z^{(\text{exp})}_{23}(\omega)}{R_H}=\frac{1}{1-\mi 
        R_HC_0\omega}\,\frac{B_{LR}(\omega)}{B_{RR}(\omega)}\,.
\end{equation}
Note that, at low frequency ($\omega R_HC_O\ll 1$), 
the result only depends
on the finite frequency properties of the $B$ quantum Hall bar {\it
i.e.}
the dipole under test in this setting.

\subsubsection{Effect of the finite frequency deviation of
$\Re(Z_{xy}(\omega))$ from $R_H$ on $Z^{(\text{exp})}_{23}$}
\label{appendix/emittances/3-contacts/impedances}

It is then useful to relate $Z_{2,3}^{(exp)}$ to the impedance of the
Hall bar $B$, which we know has a real part $\sim  R_H$, with a small deviation
for $\omega \neq 0$, as observed by metrologists studying the quantum Hall
resistance a low frequencies [15-17].

This is achieved using the relation \cite{Safi-1999-1,Degio-2010-1} between the quantum Hall bar finite
admittance matrix and its edge-magnetoplasmon scattering matrix to
obtain the following expression, valid when the left and right moving
edges of the quantum Hall bar are in total mutual influence:
\begin{equation}
        \frac{Z^{(\text{exp})}_{23}(\omega)}{R_H}=\frac{z_H(\omega)-1}{1-\mi
    R_HC_0\omega}\,
\end{equation}
in which $z_H(\omega)$ denotes the dimensionless impedance of the
quantum Hall bar $B$ in units of $R_H$. Note that
within the framework of the edge-magnetoplasmon
scattering model presented in Sec. \ref{appendix/plasmons}, 
$z_H(\omega)=2Z(\omega)/R_H$ where $Z(\omega)$ is given in terms of the
edge-magnetoplasmon scattering matrix by
Eq. \eqref{eq/impedances/Z}. It thus predicts
predicts $\Re(z_H(\omega))=1$ and therefoe
$\Re\left(Z_{23}^{(\text{exp})}(\omega)\right)=0$. 

However, we will now show that a non 
vanishing
$r(\omega)=\Re(z_H(\omega))-1$ is responsible for the non-linearity
of $X(f)$
seen on Fig. 2 of the paper.
Such a feature has been observed for long time by metrologists studying
quantum Hall resistance at low frequencies
\cite{Chua-1999-1,Schurr-2005-1,Jeckelmann-2001-1}.
In the metrology community, 
this finite frequency deviation $r(\omega)$ from the dc
quantum Hall value 
is commonly attributed to parasitic effects such as self and resistance of 
ohmic contacts and bond wires and thus depend 
on configurations and samples. A quadratic dependence on the frequency is usually found
in the form $r(\omega)=a_2(\omega/2\pi)^2$. 
The $a_2$ parameter has been measured in many configurations and
samples and has been found to be
positive or negative but is always around (or below)
$\SI{e-7}{\per\kilo\hertz^2}$ 
\cite{Chua-1999-1,Jeckelmann-2001-1,Schurr-2005-1}. 

To discuss the experimental results, which include both the real and
imaginary parts of $Z_{23}^{(\text{exp})}(\omega)$, let us therefore write
\begin{equation}
z_H(\omega)=1+r(\omega)+\mi \omega \tau_H(\omega)
\end{equation}
in which $r(\omega)$ denotes the frequency-dependent deviation to $R_H$
(when non zero),
and $\tau_H(\omega)=L_{\text{eff}}(\omega)/R_H$ is the dimensionless frequency-dependent
effective $RL$-time of the quantum Hall bar ($L=L_{\text{eff}}(\omega=0)$). The low frequency
expansion of the experimentally measured impedance
$Z^{(\text{exp})}_{23}(\omega)/R_H$ is then
\begin{subequations}
        \begin{align}
                \frac{Z^{(\text{exp})}_{23}(\omega)}{R_H} &\simeq
			r(\omega)-\omega^2 L_{\text{eff}}(\omega)C_0\\
			&+ \mi \omega\left[\frac{L_{\text{eff}}(\omega)}{R_H}+R_H
        C_0r(\omega)\right]+\mathcal{O}(\omega^3)
      \end{align}
\end{subequations}
In our experiments, the reactance
$\Im\left(Z^{(\text{exp})}_{23}(\omega)\right)$ 
is obviously dominated by the kinetic inductance
$L$:
\begin{equation}
        \Im\left(Z^{(\text{exp})}_{23}(\omega)\right)=
        \omega L\,,
\end{equation}
but $r(\omega)$ is seen through the deviations of the imaginary
part from linearity (see Fig.~2 of the paper). It is nevertheless small 
and therefore, the dominant contribution to the real 
part $\Re\left(Z^{(\text{exp})}_{23}(\omega)\right)$ is
a negative quadratic one directly proportional to the cable
capacitance:
$-\omega^2LC_0$, as observed in the inset of Fig.~2 of the paper.

Both real and imaginary part of admittance seen in Fig. 2 
suggest a quadratic behavior for $r(\omega)$.
To second order in $\omega R_KC_0$:
\begin{equation}
        \Re\left(\frac{Z^{(\text{exp})}_{23}(\omega)}{R_H}\right)
        =(R_HC_0\omega)^2\left[r_2-\frac{L/R_H}{R_HC_0}\right]
\end{equation}
where $r(\omega)\simeq r_2\,(\omega R_HC_0)^2$. 

In our experiments, the cable capacitance is $C_0=\SI{218}{\pico\farad}$. 
We deduce from Fig.~2 of the paper that, at $\nu=2$ and
$\omega/2\pi=\SI{100}{\kilo\hertz}$,
$\omega r(\omega)R_H^2C_0 \approx \SI{2.2}{\ohm}$. 
This gives $r(\omega)\approx 10^{-4}$
at $\SI{100}{\kilo\hertz}$ thereby corresponding to 
$a_2=4\pi^2r_2(R_HC_0)^2 \approx \SI{e-8}{\per\kilo\hertz^2}$, a 
value in total agreement with the deviations reported in the litterature.
Consequently, the observed deviations from pure inductive behavior can
be explained by the 
combined effect
of the cable capacitance and of the deviations of $\Re(Z_H(\omega))$ from
its dc value $R_H$.

\subsection{Four contact geometry (two on the same side)}
\label{appendix/emittances/4contacts-noleaks}

\begin{figure}[!h]
    \centering
    \includegraphics[width=8cm]{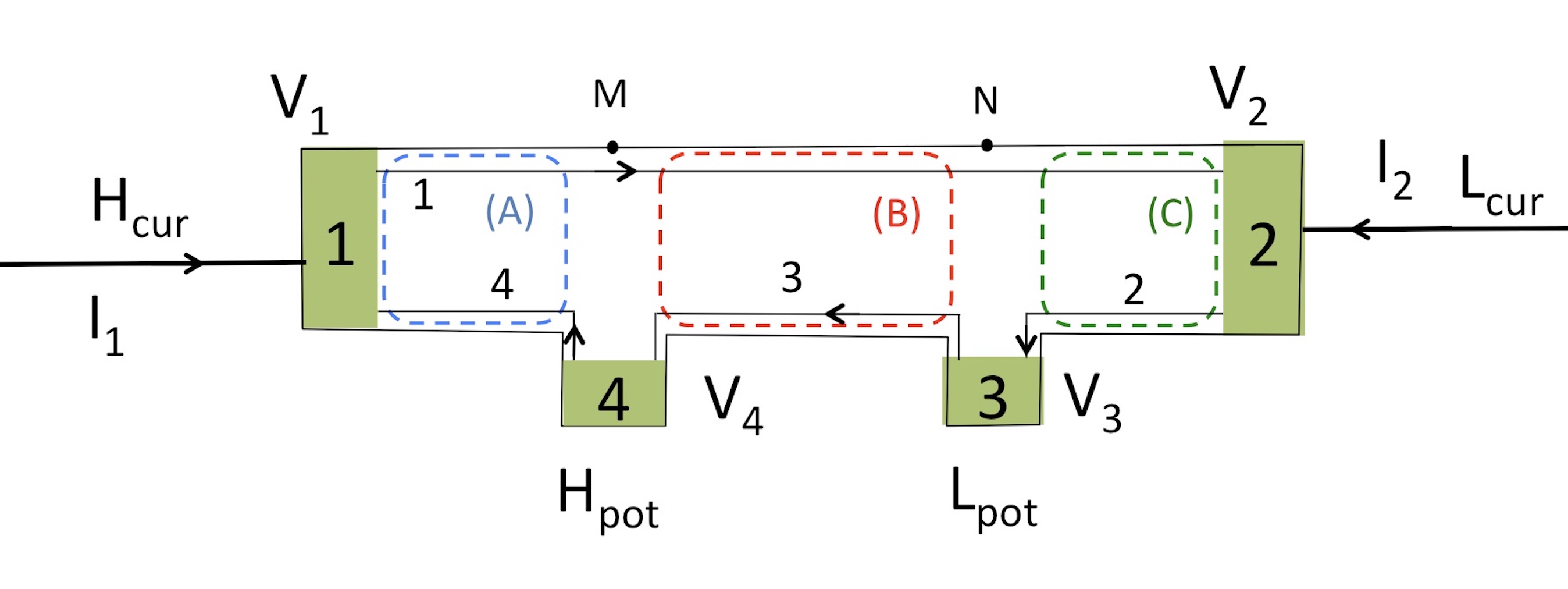}
	\caption{\label{fig:4contacts-noleaks} The device under
	test has now four contacts connected to sample holder, two of them
	being on same side. The quantum Hall bar is divided into three parts, shown as colored box, 
	A (blue), B(red) and C(green). $H_{\text{pot}}$ and $L_{\text{pot}}$ are connected
	respectively to contacts 4 and 3. In this configuration, the low
	frequency impedance depends only on the central Hall bar B. }
\end{figure}

Let us now turn to the case of a four contact sample depicted on Fig.
\ref{fig:4contacts-noleaks}. This geometry represent a four contact
measurement of the central part $B$ of the device.
Here, the parts $A$ and $C$ of the device play the role of the leads
that connect $B$ to the current injection reservoirs which are the Ohmic
contacts $1$ and $2$. The Ohmic contacts $3$ and $4$ are used for
measuring the voltage difference between the two port of the 
would-be dipole $B$. 

In this section, the capacitances $C_0$ of the cables will be omitted for
simplicity. This does not change the physics of the problem but greatly
simplifies the computations. Nonetheless, we know from Eq.
\eqref{eq-Ohmic-input-output} that the same
prefactor $(1-\mi R_HC_0 \omega)^{-1}$ will appear in the results when capacitances
are taken into account.

We assume that $A$, $B$ and $C$ can be modeled as quantum Hall bars
characterized by their edge-magnetoplasmon scattering matrices.
Exactly as before, this amounts to neglecting 
Coulomb interactions outside the Hall bars themselves, an hypothesis
consistent with the total screening hypothesis stated in Sec.
\ref{appendix/plasmons/model}. 
Let us denote by $M$ and $N$ the points on edge channel 1, which
respectively face contacts 4 and 3. Introducing the EMP scattering
matrices $\mathbf{A}$, $\mathbf{B}$ and $\mathbf{C}$ respectively
associated with the three quantum Hall bars 
$A$, $B$ and 
$C$ represented on Fig. \ref{fig:4contacts-noleaks}, the dynamics of the
three quantum Hall bars is described by:
% Corrigée!!!
\begin{subequations}
	\begin{align}
\begin{pmatrix}
i_M\\
	i_{4,\text{out}}
\end{pmatrix} &=A
\begin{pmatrix}
	i_{1,\text{in}}\\
i_{4,in}
\end{pmatrix}
\label{matrixA}\\
\begin{pmatrix}
i_N\\
	i_{3,\text{out}}
\end{pmatrix} &=B
\begin{pmatrix}
i_{M}\\
	i_{3,\text{in}}
\end{pmatrix}
\label{matrixB}\\
\begin{pmatrix}
	i_{1, \text{out}}\\
	i_{2,\text{out}}
\end{pmatrix} &=C
\begin{pmatrix}
i_{N}\\
i_{2,in}
\end{pmatrix}
\label{matrixC}
	\end{align}
\end{subequations}
in which, for compactness, the $\omega$ dependence is not recalled.
In the experimental configuration depicted in Fig.
\ref{fig:4contacts-noleaks}, $V_3=0$ and we measure $V_4$. 
We calculate $I_2$ and $V_4$, in order to find $Z_{42}^{(\text{exp})}=V_4/I_2$. The lead
potentials fix the incoming currents injected by the reservoirs.
Moreover, because of the
high impedance of the voltmeter connected to $H_{\mathrm{pot}}$ and $L_{\mathrm{pot}}$, we neglect
the currents leaking into these Ohmic contacts.
We thus have:
% Checked.
\begin{subequations}
\begin{align}
	i_{1,\text{in}}&=G_HV_1,\quad i_{2,\text{in}}=G_HV_2\\
	i_{3,\text{in}}&=0,\quad i_{4,\text{in}}=G_HV_4\\
	i_{3,\text{in}}&=i_{2,\text{out}},\quad
	i_{4,\text{in}}=i_{3,\text{out}}\\
	I_2&=i_{2,\text{in}}-i_{1,\text{out}}.
\end{align}
\end{subequations}
From Eqs. \eqref{matrixA} and \eqref{matrixB}, we obtain :
% Checked.
\begin{align}
	G_HV_4&=i_{4,\text{in}}=i_{3,\text{out}}\nonumber\\
	i_{3,\text{out}}&=
	B_{LR}(A_{RR}G_HV_1+A_{RL}G_HV_4)
\end{align}
or, equivalently:
% Checked.
\begin{equation}
V_4=\frac{B_{LR}A_{RR}}{1-B_{LR}A_{RL}}\,V_1\,.
\label{V4}
\end{equation}
On the other hand, calculating $I_2$:
% Checked.
\begin{equation}
	I_2=G_HV_2-i_{2,\text{out}}=G_HV_2(1-C_{RL})-C_{RR}i_N \, ,
\end{equation}
using now  Eq. \eqref{matrixC}, this becomes a current division relation
$I_2=\mathcal{C}\,i_N$
where 
\begin{equation}
\mathcal{C}=\frac{-C_{RR}C_{LL}-C_{LR}+C_{LR}C_{RL}}{C_{LL}} \,.
\label{C}
\end{equation}
Eqs. \eqref{matrixA} to \eqref{matrixB} then shows that $i_N=B_{RR}i_M$. We now use
Eq. \eqref{matrixA} to express $i_M$ in terms of $G_HV_1$ and $G_HV_4$. 
Using Eq. \eqref{V4}, we obtain :
\begin{equation}
	i_N=
	\left( \frac{B_{RR}A_{RR}}{1-B_{LR}A_{RL}}\right)G_HV_1\,.
\end{equation}
This gives us the total current entering lead $2$ in terms of the
applied voltage $V_1$ at the right Ohmic contact:
\begin{equation}
I_2=\mathcal{C}\frac{B_{RR}A_{RR}}{1-B_{LR}A_{RL}}G_HV_1
\end{equation}
Using one last time the voltage division relation \eqref{V4}, 
the dimensionless impedance $z_{24}=Z_{24}^{(\text{exp})}/R_H=G_HV_4/I_2$ is thus finally given by:
\begin{equation}
	z_{24}=\frac{B_{LR}}{B_{RR}\times \mathcal{C}}\,.
\label{zH4C}
\end{equation}
As in the three contact case, the fraction $B_{LR}/B_{RR}$ appears in the results
showing that the four point measurement does gives us information on 
the $B$ quantum Hall bar.  However, the matrix $C$
is involved via the quantity
$\mathcal{C}$. 
Nevertheless, it is possible to show that, to the lowest order frequency
expansion, $z_{24}$ does not depend on the physical relevant information
contained within the C matrix.  For this we once again 
use the low frequency expansion of the $C$
scattering matrix (see Eq. \eqref{eq/scattering/totally-screened}):
\begin{align}
	C_{LL}&=C_{RR}=1+C_{RL} \\
	C_{LR}&=C_{RL}=\frac{-\mi\omega l_C}{v_{\text{d}}(2+\nu \alpha_{\nu})}=-\mi\omega \tau_C
\end{align}
where $l_C$ is the length of Hall bar $C$, and $\tau_C=\frac{l_C}{2v_{\text{d}}(1+\nu\alpha_{\nu}/2)}$ 
is the transit time across the  Hall bar $C$.
These formulas allow the calculation of quantity $\mathcal{C}$ from its 
definition by Eq. \eqref{C} at the first order in $\omega\tau_C$:
\begin{equation}
	\mathcal{C}=-1+2\mi\omega \tau_C+o\left((\omega\tau_C)^2\right)\,.
\end{equation}
Finally, using Eq. \eqref{zH4C},
\begin{equation}
	z_{24}=\frac{-\mi\omega \tau_B}{(1-i\omega \tau_B)(-1+2\mi\omega
	\tau_C)}\simeq 
	\mi\omega \tau_B +o\left((\omega\tau_C)^2\right)\,
\end{equation}
showing that the measured inductance corresponds to the transit time in
Hall bar $B$. Thus, the length
involved in the determination of velocities is $l=l_B$.

\section{Discrete element description from edge-magnetoplasmon
scattering}
\label{appendix/plasmons}

In this Section, we present the computation of the EMP scattering matrix
for an ungated quantum Hall bar and derive the corresponding description
in terms of discrete elements valid at low frequency.

\subsection{The model}
\label{appendix/plasmons/model}

\begin{figure}
    \centering
	\includegraphics[width=8cm]{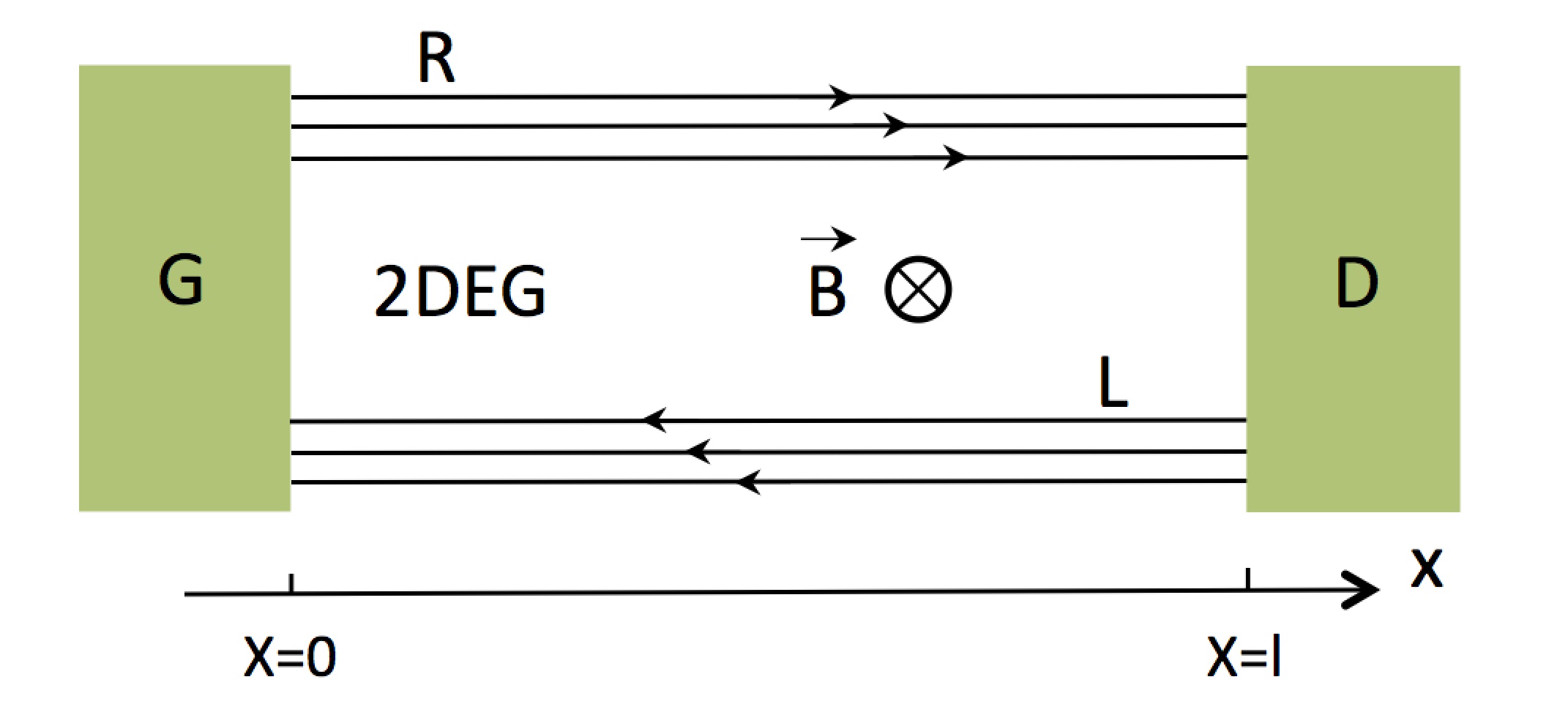}
    \caption{\label{fig/QH-bar} A simple quantum Hall bar at filling fraction
    $\nu=3$ connecting the $G$ lead to the $D$ lead. Along the $R$ edge
    channels, electrons are right movers flowing from lead $G$ to lead
    $D$ whereas for edge channels $L$, they are left movers.}
\end{figure}

We consider a quantum Hall bar of length $l$ in the integer quantum Hall regime with filling
fraction $\nu\geq 1$ as depicted on Fig. \ref{fig/QH-bar}. We denote by $R$ (resp. $L$) 
the upper (resp. lower) edge channels which are
right (resp. left) movers as depicted on Fig.
\ref{fig/QH-bar}. We denote by
$j^{(\text{in})}_\alpha$ the incoming current entering the Hall bar on
side $\alpha=R$, $L$ and $j^{(\text{out})}_\alpha$ the outgoing one. 
Our goal is to relate the outgoing currents to the incoming ones in the
presence of Coulomb interactions. 

Here $v_{\text{d}}$ denotes the charge transport
velocity, which for simplicity we take identical for all chiral edge channels.
Coulomb interactions are described by assuming a discrete element model
{\it à la} Christen-Büttiker \cite{Christen-1996-1} in which all the electrons within channels of the
same chirality see the same time dependent potential $U_{R}(t)$ (for
right movers) and $U_L(t)$ (for left movers). These potentials are related
to the total charge stored within the $L$ and $R$ edge channels by a capacitance matrix
\begin{equation}
\label{eq/Coulomb}
	\mathbf{Q}=\mathbf{C}\,\mathbf{U}\,
\end{equation}
in which 
\begin{equation}
	\mathbf{Q}=
	\begin{pmatrix}
Q_R\\ 
Q_L 
\end{pmatrix},
	\quad \mathbf{U}=
	   \begin{pmatrix}
U_R\\
U_L
\end{pmatrix}
\end{equation}
and $\mathbf{C}$ denotes the capacitance matrix
\begin{equation}
	\label{eq/capacitance-matrix}
	\mathbf{C}=
    \begin{pmatrix}
        C_H & -\eta\,C_H \\
        -\eta\,C_H  & C_H
    \end{pmatrix}\,,
\end{equation}
where $0\leq \eta\leq 1$ depending on the screening of Coulomb
interaction between channels of opposite chiralities by nearby gates and
other on-chip conductors. Eq. \eqref{eq/Coulomb} plays the role
of a solution
of the Poisson equation relating the electrical potential to the charge
density. In this work, unless otherwise stated, we assume that the left and right chiral
edge channels are in total influence and therefore that $\eta=1$. The
capacitance $C_H$ 
can then be computed as the geometric capacitance of a stripline 
capacitor (see Sec. \ref{appendix/capacitance}).

The charges within the edge channels are directly related to the incoming
and outgoing currents through charge conservation which, in the Fourier
domain and in vector notation, is written as
\begin{equation}
	\label{eq/charge-conservation}
	\mathbf{j}^{(\text{in})}(\omega) -
	\mathbf{j}^{(\text{out})}(\omega) = -\mi \omega
	\mathbf{Q}(\omega)\,. 
\end{equation}
The last equation we need is the equation of motion for the
charge or current density within the Hall bar \cite{Cabart-2018-1}:
% OK checked.
\begin{equation}
	\label{eq/equation-motion}
	\mathbf{Q}(\omega) + C_q(l,\nu;\omega)\mathbf{U}(\omega)=
	\frac{l}{v_{\text{d}}}\,
	f\left(\frac{\omega l}{v_{\text{d}}}\right)\,
	\mathbf{j}^{(\text{in})}(\omega)\,,
\end{equation}
in which 
$f(x)=(\me^{\mi x}-1)/\mi x$, and
\begin{equation}
	\label{eq/effective-Cq}
	C_q(l,\nu;\omega)=\frac{\nu e^2l}{hv_{\text{d}}}\,f\left(\frac{\omega
	l}{v_{\text{d}}}\right)
\end{equation}
represents the effective quantum capacitance at frequency $\omega/2\pi$ 
for the edge channels of
length $l$ and filling fraction $\nu$. Eq. \eqref{eq/equation-motion}
expresses that the charge stored within each edge channels
comes
from what is injected by the reservoirs and from the shift of the bottom
of the electronic bands by the electric potential seen by all electrons. 

\subsection{Edge magnetoplasmon scattering}
\label{appendix/plasmon/scattering}

Using Eqs. \eqref{eq/Coulomb} and \eqref{eq/charge-conservation} to
eliminate 
$\mathbf{Q}(\omega)$ and 
$\mathbf{U}(\omega)$ from Eq. \eqref{eq/equation-motion} leads to
\begin{align}
(C_q(l,\nu;\omega) &+\mathbf{C})\,
\mathbf{j}^{(\text{out})}(\omega) =\nonumber \\ 
& \left(C_q(l,\nu;\omega)+
\me^{\mi \omega l/v_{\text{d}}}\mathbf{C}
\right)\,
\mathbf{j}^{(\text{in})}(\omega)\,.
\label{eq/in-out}
\end{align}
Solving this equation leads to the scattering matrix $S(\omega)$
relating incoming to outgoing electrical currents at $\omega$. From
the bosonization point of view, this is
the scattering matrix for the bosonic counter-propagating edge-magnetoplasmon modes 
carrying
the total charges of the $\nu$ right and $\nu$ left moving edge
channels.

This scattering matrix 
depends on a dimensionless coupling constant $\alpha_\nu$
characterizing the strength of Coulomb interactions within each edge
channel by the way of the ratio of the bare quantum capacitance of a single
edge channel with drift velocity $v_{\text{d}}$ to the
geometric capacitance~\footnote{The $l$ dependence of $C_H$ is restored
here to stress that we are in the limit of a long quantum Hall bar
($l\gg W$) where it is linear in $l$.}:
\begin{equation}
	\alpha_\nu=\frac{C_q(l,\nu;\omega=0)}{C_H(l)\nu}=\frac{e^2
	l/hv_{\text{d}}(\nu)}{C_H(l)}\,.
\end{equation}
The higher this number, the higher is Coulomb's energy 
$e^2/C_H(l)$ with respect to the kinetic energy scale
$hv_{\text{d}}/l$. The geometric capacitance can be evaluated using standard
electrostatics as explained in Sec. \ref{appendix/capacitance}. The
result comes under the form
\begin{equation}
	C_H(l)=\varepsilon_0\varepsilon_rl\,f_{\text{bar}}(W,W_H(\nu))
\end{equation}
in which $\varepsilon_r$ is the relative permittivity of the material
and $f_{\text{bar}}(W,W_H(\nu))$ is a geometrical factor that depends on the width
$W$ of the quantum Hall bar and on $W_H(\nu)$, the width of the
system of $\nu$ copropagating edge channels at the edge of the quantum
Hall bar. The
latter is proportional to the filling fraction $\nu$ and, in the
$W_H(\nu)\ll W$ limit, the geometric
factor
$f_{\text{bar}}$ depends logarithmically on the aspect ratio $W/W_H(\nu)$ 
as shown in Sec. \ref{appendix/capacitance}. 
The dimensionless coupling constant 
\begin{equation}
	\label{eq/single-channel-coupling}
	\alpha_\nu=\frac{2\,\alpha_{\text{eff}}(\nu)}{f_{\text{bar}}(W,W_H(\nu))}\,
\end{equation}
is therefore proportional to the
effective fine structure constant at filling fraction $\nu$
\begin{equation}
	\label{eq/effective-fine-structure-constant}
	\alpha_{\text{eff}}(\nu)=\frac{e^2}{4\pi\varepsilon_r\varepsilon_0\hbar
	v_{\text{d}}(\nu)}=
	\frac{\alpha_{\text{qed}}}{\varepsilon_r}\,\frac{c}{v_{\text{d}}(\nu)}\,.
\end{equation}
up to a subdominant
logarithmic dependence in $\nu$ arising from classical electrostatics
(see Sec. \ref{appendix/capacitance}).
The $\omega$ dependence is an $X=\omega l/v_{\text{d}}$ dependence involving
the free electron time of flight $l/v_{\text{d}}$ along the edges of the quantum Hall
bar. 
Solving Eq. \eqref{eq/in-out} leads to the edge magnetoplasmon
scattering amplitudes
\begin{subequations}
\label{eq/scattering/totally-screened}	
	\begin{align}
	\label{eq/scattering/totally-screened/off-diagonal}
		S_{RL}(\omega) &=S_{LR}(\omega)= 
		\frac{-\mi Xf(X)}{2+\nu\alpha_\nu f(X)}\\ 
	\label{eq/scattering/totally-screened/diagonal}
		S_{RR}(\omega)&=S_{LL}(\omega)=1-S_{RL}(\omega)\,.
	\end{align}
\end{subequations}
Note that the relation between diagonal and off diagonal $S$ matrix elements given by Eq. 
\eqref{eq/scattering/totally-screened/diagonal} is only valid for total
screening ($\eta = 1$).
One should also
note that
because of the symmetric setup considered here,
the edge magnetoplasmon scattering matrix is determined by its diagonal coefficient
$S_{\text{d}}(\omega)=S_{RR}(\omega)=S_{LL}(\omega)$ 
and its off-diagonal coefficient
$S_{\text{od}}(\omega)=S_{LR}(\omega)=S_{RL}(\omega)$. They
satisfy the unitarity conditions
\begin{subequations}
	\label{eq/unitarity}
	\begin{align}
		|S_\mathrm{d}(\omega)|^2+|S_{\mathrm{od}}(\omega)|^2&=1\\
		\Re(S_{\mathrm{d}}(\omega)S_{\mathrm{od}}(\omega)^*)&=0\,.
	\end{align}
\end{subequations}
which ensure energy conservation within the quantum Hall bar.

\subsection{Finite frequency impedances}
\label{appendix/plasmons/admittances}

\begin{figure}
    \centering
	\includegraphics[width=8cm]{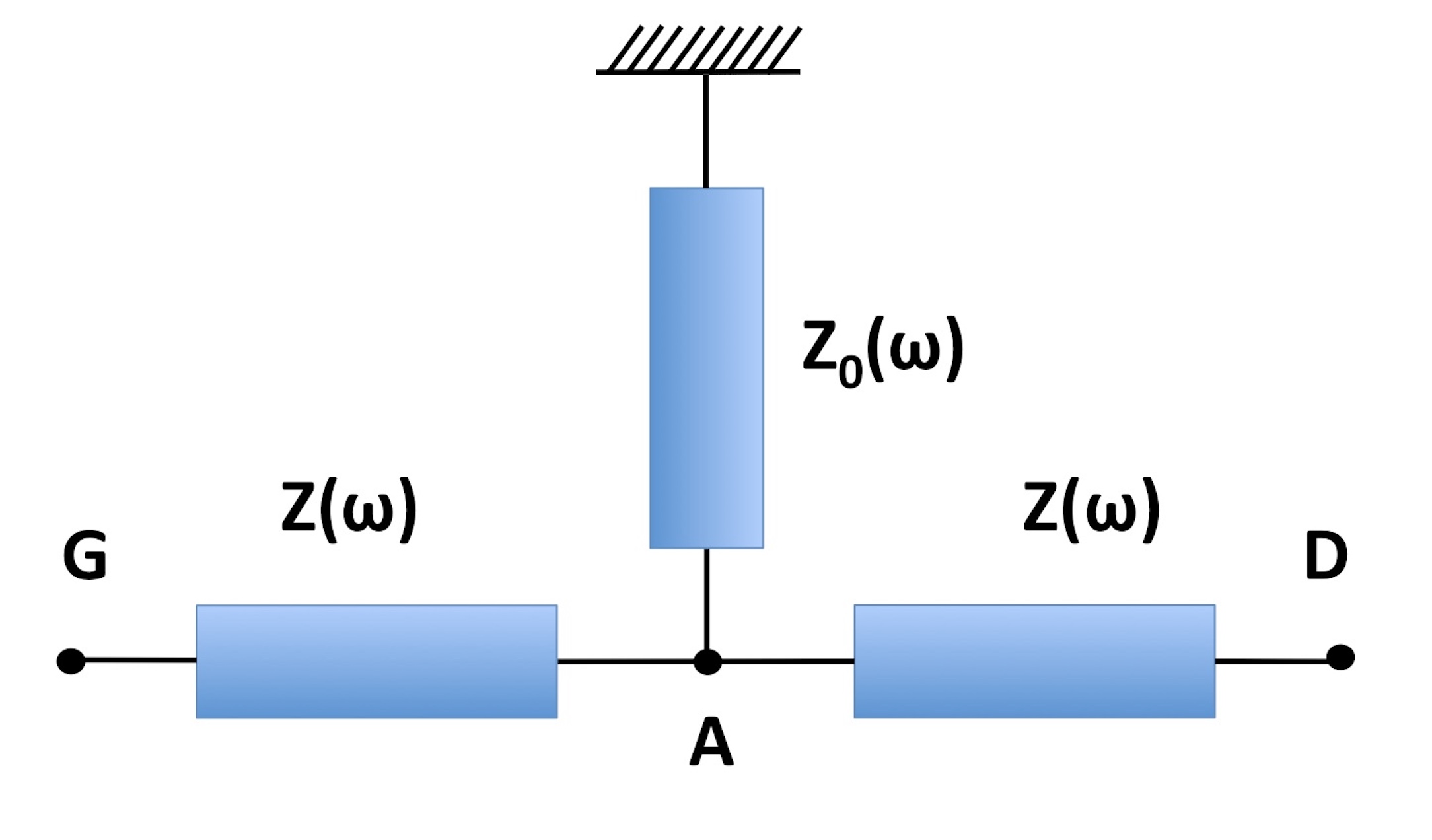}
    \caption{\label{fig/QH-circuit} The lumped element circuit
    equivalent to simple quantum Hall bar depicted on Fig.
    \ref{fig/QH-bar}. The $0$ lead corresponds to metallic gates and
    conductors to which the edge channels are capacitively coupled in
    the case of partial mutual capacitive coupling between the $R$ and
    $L$ channels ($\eta<1$).
    We expect $\Re(Z(\omega))=R_H/2$ and
    $\Im(Z(\omega))$ to be inductive at low frequency whereas
    $\Re(Z_0(\omega))=0$ and $Z_0(\omega)$ is expected to be capacitive
	in general, and to be infinite in the totally screened case ($\eta=1$).
    }
\end{figure}

In the spirit of Büttiker, let us derive a discrete element circuit
description of the quantum Hall bar. In full generality, this description
involves a
three terminal circuit in order to take into account the electrostatic
coupling to external gates which are assumed to be at the ground here
(see Fig. \ref{fig/QH-circuit}). In our experimental, ungated samples,
this coupling is expected to be small as we shall see.
Note that the symmetry of the Hall bar
under magnetic field reversal
justifies considering two identical impedances $Z(\omega)$ on both sides
of the central node $A$ on Fig. \ref{fig/QH-circuit}. 
For this tri-terminal circuit, one can combine the relation between the
potentials $V_G$, $V_D$, $V_A$ and $V_0=0$ (by gauge invariance) and the
current sum rule $I_G+I_D+I_0=0$
to relate
$I_G$ and $I_D$ to $V_G$ and $V_D$: 
\begin{equation}
	\label{eq/circuit/Z-equations}
	\begin{pmatrix}
		Z+Z_0 & Z_0 \\
		Z_0 & Z+Z_0
	\end{pmatrix}\cdot
	\begin{pmatrix}
		I_G\\
		I_D
	\end{pmatrix}
	=
	\begin{pmatrix}
        V_G\\
        V_D
    \end{pmatrix}\,.
\end{equation}
On the other
hand, the finite frequency admittance of the quantum Hall bar can be
obtained from the edge magnetoplasmon scattering matrix
which, in the present case is symmetric. We can then infer expressions
for the impedances $Z(\omega)$ and $Z_0(\omega)$ in terms of the 
EMP scattering matrix
of the quantum Hall bar:
\begin{subequations}
	\begin{align}
		\label{eq/impedances/Z0}
		Z_0(\omega)&=R_H\,\frac{S_{\mathrm{d}}(\omega)}{(1-S_{\mathrm{od}}(\omega))^2-S_{\mathrm{d}}(\omega)^2}\,,\\
		\label{eq/impedances/Z}
		Z(\omega)&=\frac{R_H}{1+S_{\mathrm{d}}(\omega)-S_{\mathrm{od}}(\omega)}\,.
	\end{align}
\end{subequations}
Note that Eqs. \eqref{eq/impedances/Z0} and
\eqref{eq/impedances/Z} have been obtained without any explicit
assumption of total screening. In the case of $\eta = 1$ (total
screening, see Eq. \eqref{eq/capacitance-matrix}), $Z_0$ is expected to be infinite 
and is indeed found to be infinite since, in this case,
$S_{\text{d}}(\omega)+S_{\text{od}}(\omega)=1$ 
(see Eq. \eqref{eq/scattering/totally-screened/diagonal}).

To obtain the simplest effective circuit description, we just derive the
dominant terms of the low frequency expansion of $Z(\omega)$ and
$1/Z_0(\omega)$ by expanding the edge-magnetoplasmon scattering matrix
in powers of $\omega$. This leads to $1/Z_0(\omega)=0$ (total screening)
and
\begin{equation}
	Z(\omega)=\frac{R_H}{2}\left(1-\frac{\mi \omega
	l/v_{\text{d}}}{2+\nu\alpha_\nu}\right)
\end{equation}
which shows that the impedance $Z(\omega)$ can be viewed as the series
addition of a resistance $R_H/2=R_K/2\nu$ ($\nu$ parallel single channel
contact resistances $R_K/2$) and an inductance $L_Z(\nu)$ such that
\begin{equation}
	\frac{2L_Z(\nu)}{R_H(\nu)}=\frac{l/2\,v_{\text{d}}(\nu)}{1+\nu\alpha_\nu/2}\,.
\end{equation}
The quantum Hall bar is then a dipole with impedance $2Z(\omega)$ which
then behaves as an inductance in series with the quantum Hall resistance
$R_H$ at low frequency.
Since $2L_Z$ is the total inductance $L$ of the quantum
Hall bar,
it follows that the electronic time of flight $l/v_{\text{d}}$ appearing in Eq.
\eqref{eq:kinetic-inductance} is
renormalized by Coulomb interactions. The corresponding renormalized
velocity 
\begin{equation}
	\label{eq/renormalized-velocity}
	v_{\text{eff}}(\nu)=v_{\text{d}}(\nu)\,\left(1+\frac{\nu\alpha_\nu}{2}\right)
\end{equation}
can, in a sense, be interpreted as an effective charge velocity. 

To understand this more precisely,
let us imagine that the quantum Hall bar could be viewed as an ideal coaxial
cable with $\nu$ channels with plasmonic velocity $v_P$. This would lead
to
a diagonal plasmon scattering matrix:
$S_{\text{od}}(\omega)=0$ and $S_{\text{d}}(\omega)=\me^{\mi\omega
l/v_P}$. The corresponding admittance $Z(\omega)$  would
then be given by
\begin{equation}
	Z_{\text{tl}}(\omega)=\frac{R_H}{2}\,\frac{1}{1+\me^{\mi\omega l/v_P}}
\end{equation}
where the index $\text{tl}$ emphasizes the transmission line description
considered here. Its
low frequency expansion would correspond to the series addition of
the contact resistance $R_H/2$ with an inductance $L_Z^{(\text{tl})}$
given by
\begin{equation}
	\frac{2L_Z^{(\text{tl})}}{R_H}=\frac{l}{2\,v_P}\,.
\end{equation}
Consequently, we also
recover the analogue of Eq. \eqref{eq:kinetic-inductance} with
$v_P$ playing the role of $v_{\text{d}}$.
However, in this transmission line model, the admittance $1/Z_0$ would
not vanish, thereby showing that such a transmission line model cannot
describe the totally screened situation. 
The edge magnetoplasmon scattering matrix derived in Sec.
\ref{appendix/plasmon/scattering} shows that, in the totally screened case, 
$R_H/Z_0$ vanishes. 

\section{Geometric capacitance}
\label{appendix/capacitance}

In this section, we compute the geometric capacitance of the quantum
Hall bar using as model a capacitor built from two coplanar strips
corresponding to the two counter-propagating edge channel systems.

\subsection{Width of edge channels} 
\label{appendix/edge-channel-width}

The model of Ref. \cite{Chklovskii-1992-1} by Schklovskii {\it et al.}  describes the
structure of quantum Hall edge channels when taking into account the effect
of Coulomb interactions. It predicts the width $\lambda_H$ (also denoted by
$a_\nu$ following Ref. \cite{Chklovskii-1992-1}) 
of the incompressible channels in term of the density gradient at the
edge as well as the 
width $b_{\nu}$ of compressible channels as a function of $\lambda_H$. 

\begin{figure}[!h]
    \centering
    \includegraphics[width=8cm]{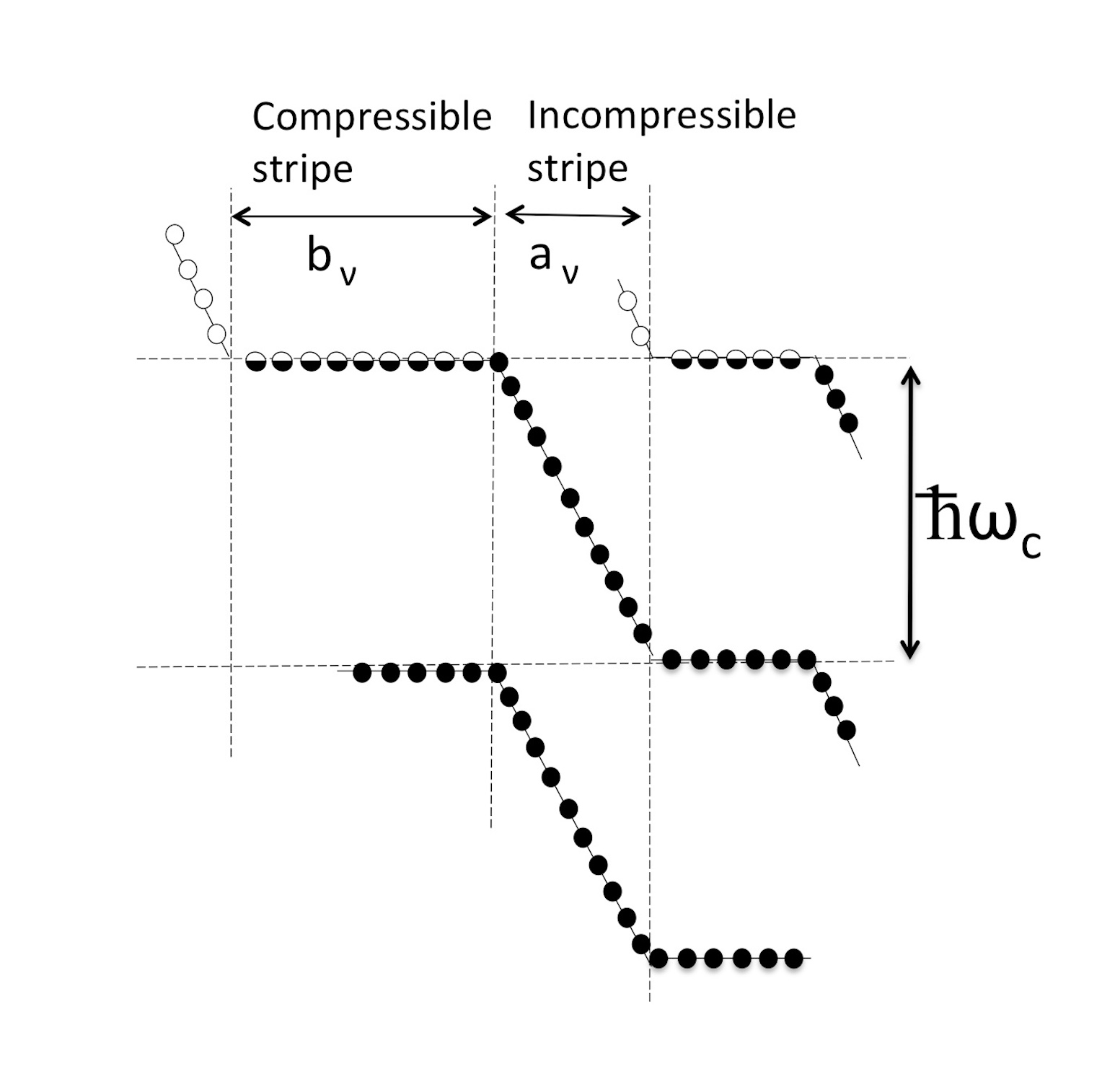}
	\caption{\label{fig:edge-channels-structure} Edge channel structure
	arising from Coulomb interaction effects in a mean field theory
	approach \cite{Chklovskii-1992-1}: incompressible stripes alternate with compressible
	stripes in which electrons are located right at the Fermi energy.}
\end{figure}

The width of incompressible stripes is obtained by identifying the
Landau gap with the energy variation of an electron in 
the confining potential $U_\nu$ (see Figure \ref{fig:edge-channels-structure}):
\begin{equation}
a_{\nu} \times e \nabla U_\nu=\hbar\frac{eB}{m^*} \;\;,
\end{equation}
where $\nabla U_\nu$ denotes the gradient of $U_\nu$ within the
incompressible part of the edge of the
sample. As explained in the main text, the confining potential may
depend on $\nu$ as discussed in
Sec.~\ref{appendix/nu-potential}. For simplicity, we shall assume that this gradient has
indeed the same dependence on $\nu$ as the charge transport velocity
$v_{\text{d}}(\nu)$ present in the equations of motion
\eqref{eq/equation-motion} and \eqref{eq/effective-Cq}. 
This
leads to:
\begin{equation}
	\label{eq/a-nu}
	a_{\nu}= \frac{\hbar}{m^*v_{\text{d}}(\nu)}\,,
\end{equation}
where we have replaced the drift velocity within the incompressible
stripe $(E_y/B_z)_\nu$ by the drift velocity along the
edge by $v_{\text{d}}(\nu)$.
Note that this expression is formally the Compton wavelength associated
with a relativistic particle of mass $m^*$ and effective speed of light
$v_{\text{d}}(\nu)$. This explains why we denote it by $\lambda_H$ in the following. This
result is not surprising, the width of the incompressible stripe is the
minimal one for
closing the cyclotron gap 
for an electronic excitation in the confinement potential,
as shown in Ref. \cite{Chklovskii-1992-1}.

Once the width of an incompressible stripe is known, the width of a
compressible stripe is given by
\begin{equation}
	\label{eq/b-nu}
b_{\nu}= \frac{\pi}{4} \frac{\lambda_H^2}{a_B}
\end{equation}
where $a_B=\hbar^2 \varepsilon / m^* e^2$ denotes the effective Bohr
radius in the material.

\subsection{On the $\nu$-dependence of $v_{\text{d}}(\nu)$}
\label{appendix/nu-potential}

In this section, we discuss the $\nu$-dependence of the confining
potential $v_{\text{d}}(\nu)$. In the regime where no compressible
stripes are present, this would be the chiral drift velocity 
associated with the confining potential at the edge of the sample 
$v_{\text{d}}=\nabla U/eB$. In the case of a smooth confining potential,
compressible stripes may appear and the velocity $v_d(\nu)$ refers to
the effective charge velocity at the edge, arising from the dynamics of
the compressible stripe coupled to the incompressible one
\cite{Han-1997-2}. As explained
before, our main hypothesis here is that this velocity can be used in
the expressions \eqref{eq/a-nu} and \eqref{eq/b-nu} of the width of
incompressible and compressible parts of the edge channels
and that it has the same expression for all of them. 

We discuss here two different forms
for $v_{\text{d}}(\nu)$, one based on
Ref.~\cite{Mikhailov-2000-1} and the other one based on
Ref.~\cite{Chaubet-1998-1}:
\begin{equation}
    \label{eq/vnu}
    v_{\text{d}}(\nu)=
    \begin{cases}
        &v_{\text{d}}/\nu\quad \text{following\ Ref.\ \cite{Mikhailov-2000-1}} \\
        &v_{\text{d}}/\sqrt{\nu}\quad \text{following\
        Ref.\ \cite{Chaubet-1998-1}}
    \end{cases}
\end{equation}
Our discussion will be based on heuristic arguments which 
only identify $v_d(\nu)$ with a drift velocity if the edge channel
does not contain any compressible stripe \cite{Balev-1998-1,Balev-2002-1}. 
However, Han and Thouless have shown in
Ref. \cite{Han-1997-2} that,
in the presence of the compressible stripe, the velocities of the eigenmodes propagating
along the edge channel 
contain a contribution from the incompressible part drift velocity. We
think that this justifies the heuristic arguments that will now be layed
down.

In Ref.~\cite{Mikhailov-2000-1}, Mikhailov gave the following form for the edge
velocity of plasmons due to the quantum confining potential:
$v_N=\omega_c/ k_F$ (Eq. 1-60). As the cyclotron frequency $\omega_c$ is
inversely proportional to the filling factor $\nu$, and $k_F=\sqrt{2\pi
N_s}$ only depends on the density of the 2DEG,
this velocity is $v_{\text{d}}(\nu)=v_{\text{d}}/\nu$ with $v_{\text{d}}\sim\SI{4e5}{\meter\second^{-1}}$ 
for our sample. 

We have used such velocity in Eq.(1) to fit the data. However, we have varied the
numerical parameter around this value to fit our data in Fig.~4. As can
be shown from Fig.~4 of the paper, the fitting values of $v_{\text{d}}$ are
between $1/4$ and $10$ times the above estimation.

In Ref.~\cite{Chaubet-1998-1}, which studies the breakdown of the IQHE,
Chaubet and Geniet
computed a confining potential by solving the problem of an
harmonic oscillator with a boundary condition on the edge. They 
showed
that, when the wave function 
equals zero at the edge, the eigen-energies are modified compared to the boundaryless problem and 
increase, giving a realist interpretation of the confining
potential. In Fig.~9 of Ref.~\cite{Chaubet-1998-1}, the
confining potential -- represented by the energy of states -- increases
nearly linearly with the distance from the edge. We thus approximate the
gradient $\nabla U_\nu$ by the ratio of the cyclotron energy to the
magnetic length: $\nabla U_\nu =\hbar \omega_c /l_B$. The physical image
is that the potential slope corresponds to an energy gain of the order
of the Landau gap on a length scale $l_B$.
This gives a drift
velocity $v_{\text{d}}(\nu)= \sqrt{\hbar e B(\nu=1)/m^*}\,\nu^{-1/2}$ which in the case
of our sample leads to $v_{\text{d}}(\nu)=v_{\text{d}}/\sqrt{\nu}$ with
$v_{\text{d}}\sim\SI{3e5}{\meter\second^{-1}}$. 

Exactly as for the previous model, we have considered $v_{\text{d}}$ as a fit
parameter that can be varied in order to 
represent all experimental situations in our experiments. 
As can
be shown from Fig.~4 of the paper, the fitting values of $v_{\text{d}}$ are
between $1/5$ and $5$ times the above estimation, a smaller range than
with the $v_{\text{d}}(\nu)=v_{\text{d}}/\nu$ model.

\subsection{Capacitance computation}

The quantum Hall
edge channel system has a width
\begin{equation}
	\label{eq/edge-channel-system-width}
	W_H(\nu)=\frac{\nu \hbar
	}{m^*v_{\text{d}}(\nu)}\left(1+\pi^2\alpha_{\mathrm{eff}}(\nu)\right)\,,
\end{equation}
which is of the order of $\SI{90}{\nano\meter}$ per edge channel in
AlGaAs/GaAs systems. In the transverse direction, the electrons are
confined within a triangular potential well. The typical extension of
the electron's wavefunction in the transverse direction is of the order
of $\SI{5}{\nano\meter}$ \cite{Fang-1966-1}, 
therefore suggesting a rather flat shape of the edge channel.

A first estimate of the geometric capacitance of the edge channel system
can thus be obtained by computing 
the geometric capacitance per unit of length of an infinite pair of
coplanar strips of width $w$ separated by a distance $d$ \cite{Iossel-1969-1}: 
\begin{equation}
	\frac{C_H(l)}{l}=\varepsilon_0\varepsilon_r\,\frac{\mathrm{K}(\sqrt{1-x^2})}{\mathrm{K}(x)}\,,
\end{equation} 
in which $\mathrm{K}(x)$ denotes the elliptic integral of the first kind
\cite{Book-GradRyz} evaluated at
$x=d/(d+2w)$ where, in our case $w=W_H(\nu)$ and $d+2w=W$ is the total
width of the quantum Hall bar. 
Using known
asymptotics for the elliptic integral, we find the geometrical
capacitance for the quantum Hall bar of width $W$ and length $l$ at filling fraction
$\nu$
\begin{equation}
	\label{eq/capacitance/striplines}
	C_H(\nu,l)\simeq\frac{\pi\varepsilon_0\varepsilon_rl}{\ln\left[\frac{W}{8W_H(\nu)}
	\right]}
	\,.
\end{equation}
On the other hand, using a more precise
self-consistant solution of the electrical potential within quantum Hall edge
channel arising from the repartition of quantum electrons, Hirai and
Komiyama \cite{Hirai-1994-1} 
have shown that the
capacitance can be obtained as the geometric capacitance of two
parallel wires:
\begin{equation}
	\label{eq/capacitance/Harai-Komiyama}
	c'_0(\nu,l)=\frac{\pi\varepsilon_0\varepsilon_rl}{\ln\left(\frac{\me\,W}{2W_H(\nu)}\right)}\,.
\end{equation}
Expressions \eqref{eq/capacitance/striplines} 
and \eqref{eq/capacitance/Harai-Komiyama} are not identical, but both are
considered in the regime where $W\gg W_H(\nu)$, that is, when they
lead to similar estimates for the capacitance.
Both expressions are of the form 
\begin{equation}
    \label{eq/capacitance/phenomeno}
    C_H(\nu,l,\gamma)=\frac{\pi\varepsilon_0\varepsilon_rl}{\ln\left(\frac{\gamma W}{W_H(\nu)}\right)}\,,
\end{equation}
where $\gamma$ is a factor that depends on the specific model used to
describe the charge repartition within the edge channels. Here $\gamma =
\me/2$ in 
Harai {\it et al.}'s estimate and $\gamma=1/8$ for the coplanar strip
capacitance model. Using Eq. \eqref{eq/capacitance/phenomeno}, the effective coupling constant is
then
\begin{equation}
	\alpha_\nu=\frac{2\alpha_{\text{eff}}(\nu)}{\pi}\,\left[
		\ln\left(\frac{\gamma\,W/\lambda_H(\nu)}{(1+\pi^2\alpha_{\text{eff}}(\nu))\nu}\right)
		\right]\,.
	\end{equation}
Using this expression into Eq. \eqref{eq/renormalized-velocity} leads to 
\begin{equation}
\label{eq/capacitance/v0/almost}
	\frac{v_{\text{eff}}}{v_{\text{d}}}=1+\frac{\nu\alpha_{\mathrm{eff}}(\nu)}{\pi}
	\ln\left(\frac{\gamma
	W/\lambda_H(\nu)}{(1+\pi^2\alpha_{\mathrm{eff}}(\nu))\nu}\right)\,
\end{equation}
for the ratio of $v_{\text{eff}}$ to the drift velocity within a chiral edge
channel. This expression can then be rewritten in terms of the
effective single edge channel width
$\xi_H(\nu)=(1+\pi^2\alpha_{\mathrm{eff}}(\nu))\lambda_H(\nu)/\gamma$: 
	\begin{equation}
\label{eq/capacitance/v0}
		\frac{v_{\text{eff}}}{v_{\text{d}}}=1+\frac{\nu\alpha_{\mathrm{eff}}(\nu)}{\pi}
		\ln\left(\frac{W/\xi_H(\nu)}{\nu}\right)\,
\end{equation}

\section{Supplementary results} 
\label{appendix/extra-results}

In this Section, we show further experimental results obtained on sample
A (the sample discussed in the paper)
and on three other samples manufactured from three different AlGaAs/GaAs heterojunctions. 
Their wafer characteristics (electron
density and mobility of the two-dimensional electron gas) are reported in table I.
Samples have been
processed on these heterojunctions using the same design and dimensions
as for sample A.
Exeperimental results obtained on these samples have been
obtained using exactly the same protocols and they are completely similar to
the results presented in the paper. 

We first present complementary results on sample A, then we present
results obtained on the other samples, which concern influence of filling
factor $\nu$ and influence of edge channel length.  

\begin{table}[!h!]
\begin{center}
\begin{tabular}{|c||c|c|c|c|}
\hline
	Wafer & $N_s$($\si{\centi\meter^{-2}}$) &
	$\mu$($\si{\meter^2\per\volt\second}$) &  $B$ (for
	$\nu=2$) & Width ($\si{\micro\meter}$)\\
\hline \hline
	A & 5.1 & 30 & \SI{10.5}{\tesla} & 400 \\
	B & 3.3 & 50 & \SI{6.8}{\tesla} & 1600 \\
	C & 4.5 & 55 & \SI{9.2}{\tesla} & 400 \\
	D & 4.3 & 42 & \SI{9}{\tesla} & 800 \\
\hline
\end{tabular}
\caption{\label{tab:table1} Sample characteristics: carrier density,
	mobility, magnetic field at 
$\nu=2$ and width of Hall bars.}
\end{center}
\end{table}

\subsection{Complementary results on sample A} 

\begin{figure}[h!]
    \centering
    \includegraphics[width=8cm]{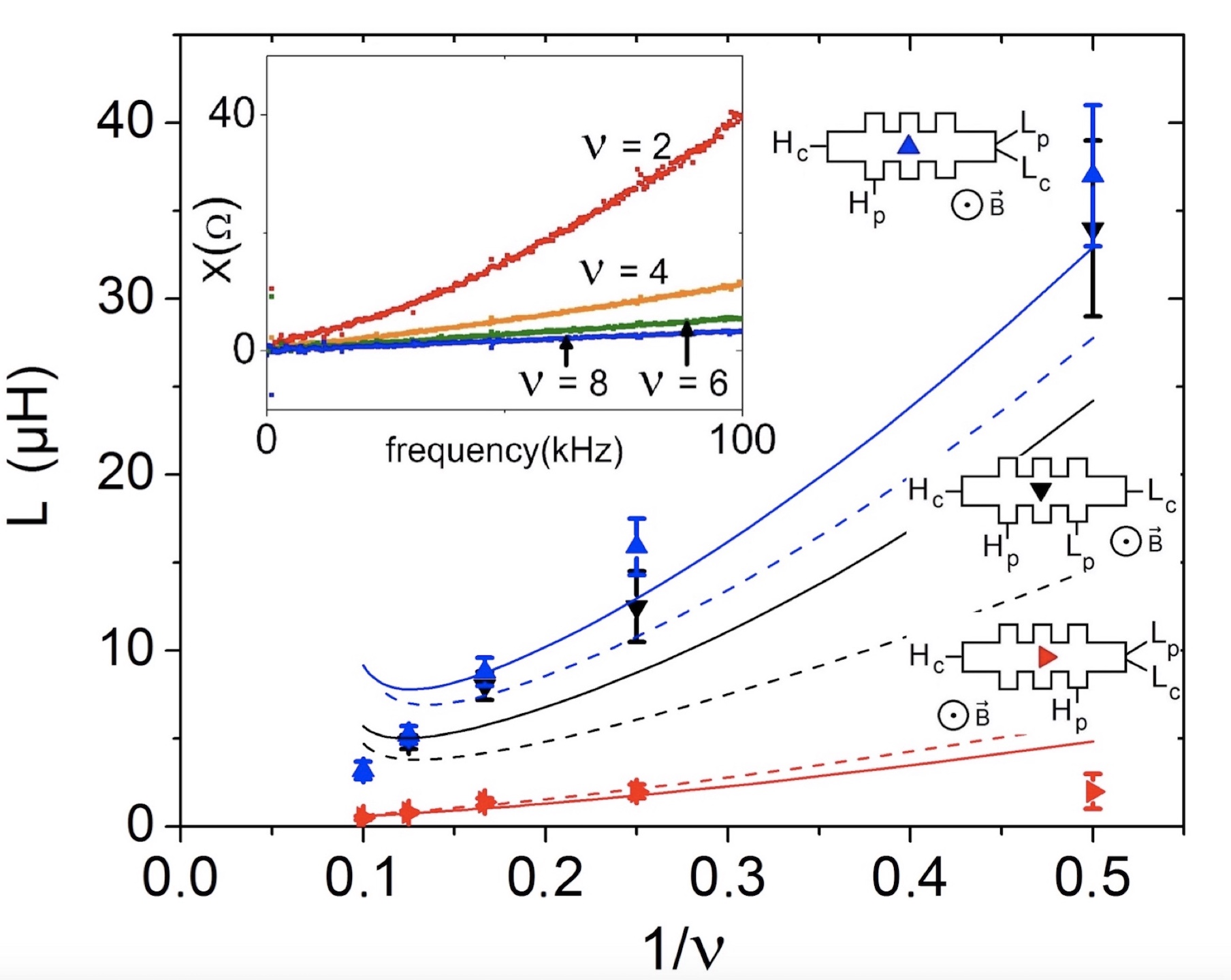}
	\caption{Kinetic inductance as a function of the inverse of filling factor for a 
	positive orientation of magnetic field. Dashed line represents the
	model $v_{\text{d}}(\nu)=v_{\text{d}}/\nu$
	with following parameters :
	$v_{\text{d}}=\SI{8e5}{\meter\second^{-1}}$ for the
	$l=\SI{1600}{\micro\meter}$ configuration, 
	$v_{\text{d}}=\SI{11e5}{\meter\second^{-1}}$ for the $l=\SI{1000}{\micro\meter}$ one,
	$v_{\text{d}}=\SI{22e5}{\meter\second^{-1}}$ for the
	$l=\SI{600}{\micro\meter}$ one. 
	In the same way, the continuous lines correspond to the
	$v_{\text{d}}(\nu)=v_{\text{d}}/\sqrt{\nu}$ model with:
   $v_{\text{d}}=\SI{1.2e5}{\meter\second^{-1}}$ for the 
    $l=\SI{1600}{\micro\meter}$ configuration,
    $v_{\text{d}}=\SI{0.5e5}{\meter\second^{-1}}$ for the $l=\SI{1000}{\micro\meter}$ one,
    $v_{\text{d}}=\SI{0.5e5}{\meter\second^{-1}}$ for the 
    $l=\SI{600}{\micro\meter}$ one.
	Note that length of edge states 
	are still the same than on Fig.~3 of the paper.}
	\label{unsurnu} 
\end{figure}

Fig.~3 of the paper shows, for a positive orientation of magnetic field, the dependence of the
quantum inertia on the inverse of the filling factor. In
Fig.~\ref{unsurnu}, these
results have been obtained for a positive orientation of magnetic field.
Once again the models $v_{\text{d}}(\nu)=v_{\text{d}}/\nu$ (dashed lines) and
$v_{\text{d}}(\nu)=v_{\text{d}}/\sqrt{\nu}$ have been considered
using a proper velocity $v_{\text{d}}$ for each configuration (see caption). 
Althougth the agreement between the experomental data and the 
theory is not as good as in the case of $B<0$ configurations, we still
see that $v_{\text{d}}(\nu)=v_{\text{d}}/\sqrt{\nu}$ better reproduces the data
(especially for the $l=\SI{1600}{\micro\meter}$ configuration. 

\subsection{Influence of filling factor for other samples} 

\begin{figure}[h!]
\centering
    \includegraphics[width=8cm]{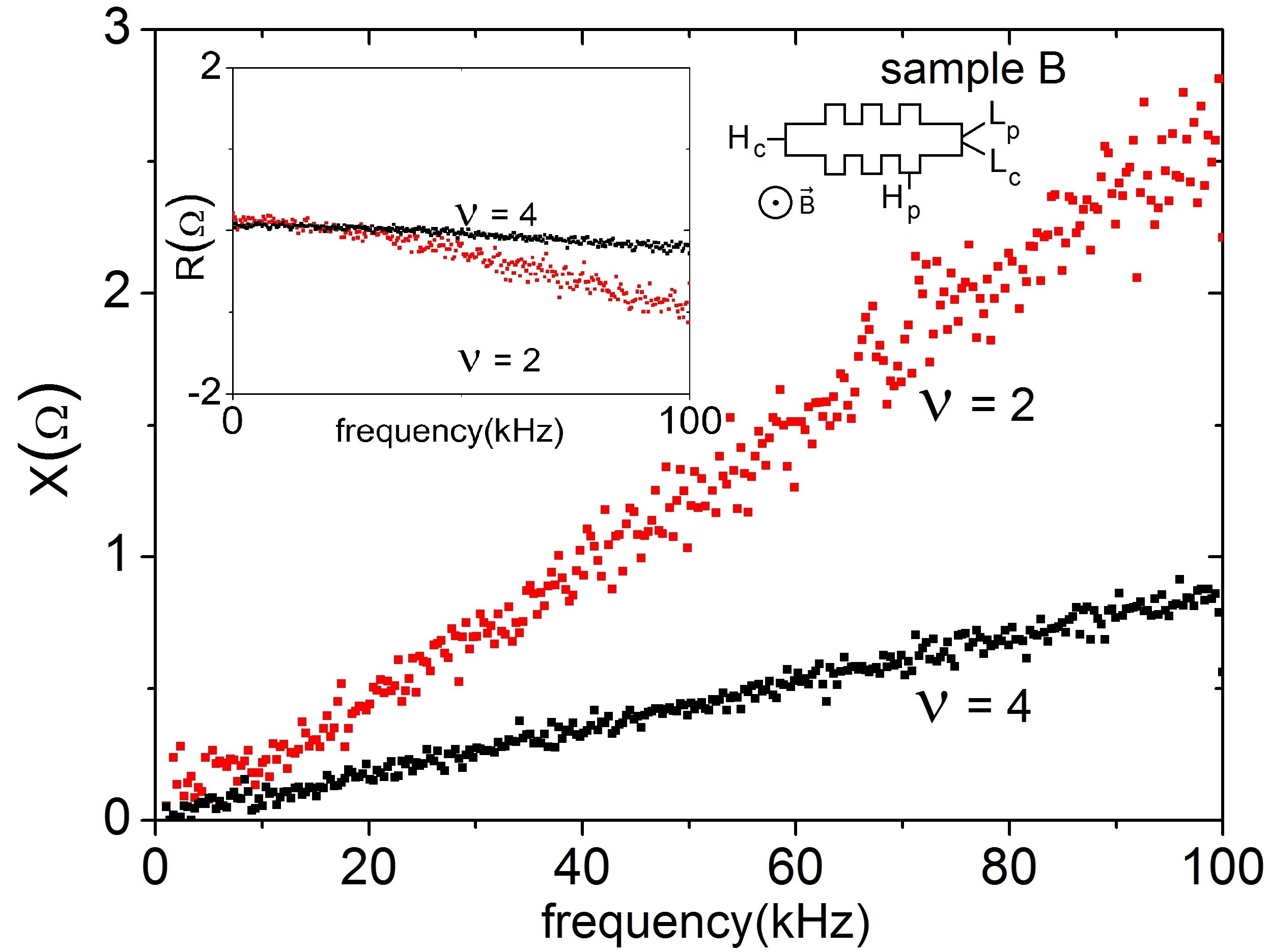}
	\caption{Reactance as a function of frequency, for
	sample B and $\nu=4$ and $4$. Configuration is shown in inset. }
	\label{FF1} 
\end{figure}

\begin{figure}[h!]
    \centering
    \includegraphics[width=8cm]{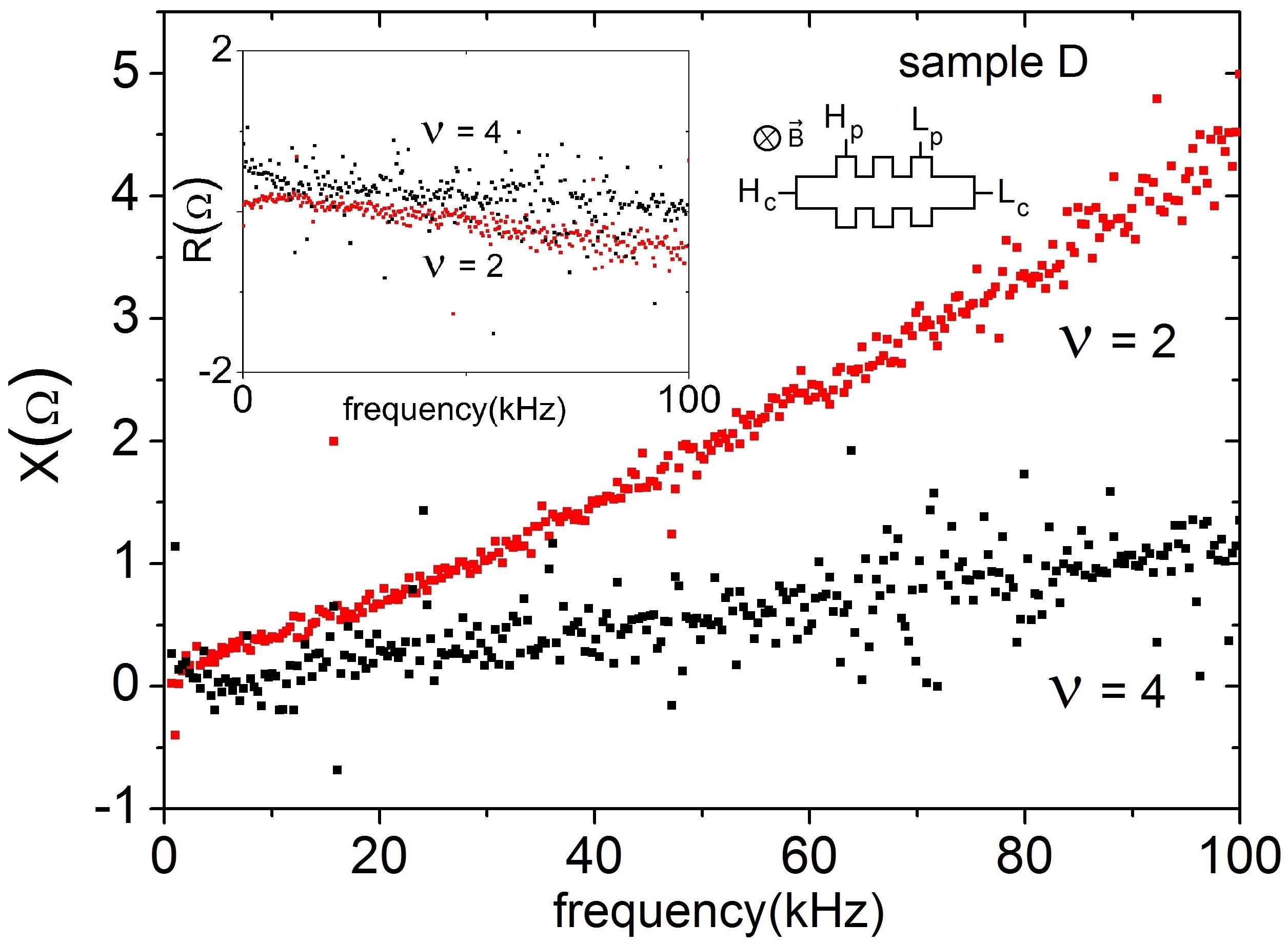}
	\caption{\label{FF3} Reactance as a function of frequency, for
	sample D and $\nu=2$ and $4$. Configuration is shown in inset. }
\end{figure}

\begin{figure}
    \centering
    \includegraphics[width=8cm]{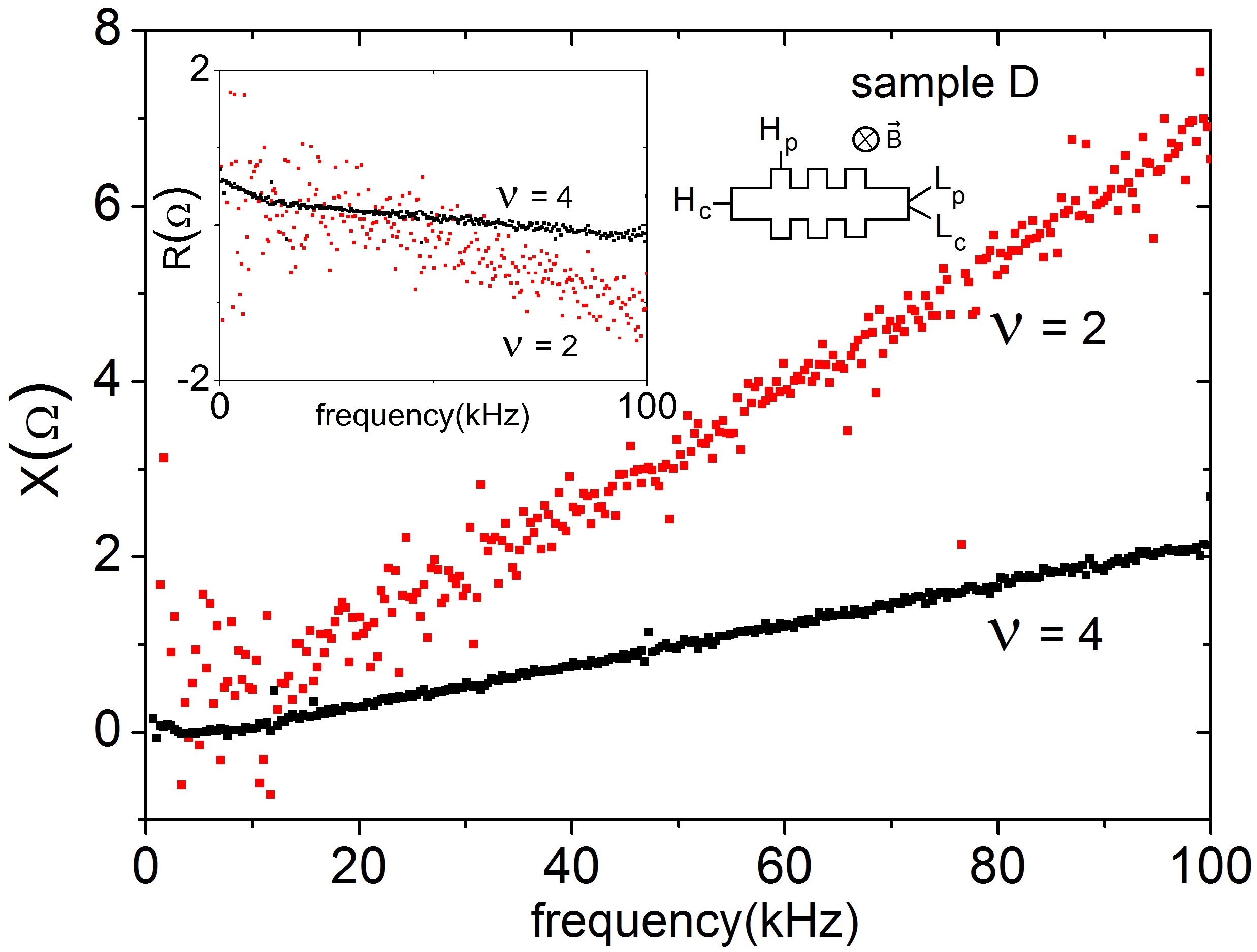}
	\caption{\label{FF4} Reactance as a function of frequency, for
	sample D at $\nu=2$ and $4$ and configuration shown in inset. }
\end{figure}

For sample A we have shown that reactance increases linearly with frequency and that
the slope depends 
strongly on the filling factor. Here, we observe the
same behavior on other samples (see Figs.~\ref{FF1} to \ref{FF4}): the slope 
decreases while the filling factor increases. 
  
\begin{figure}[h!]
    \centering
    \includegraphics[width=8cm]{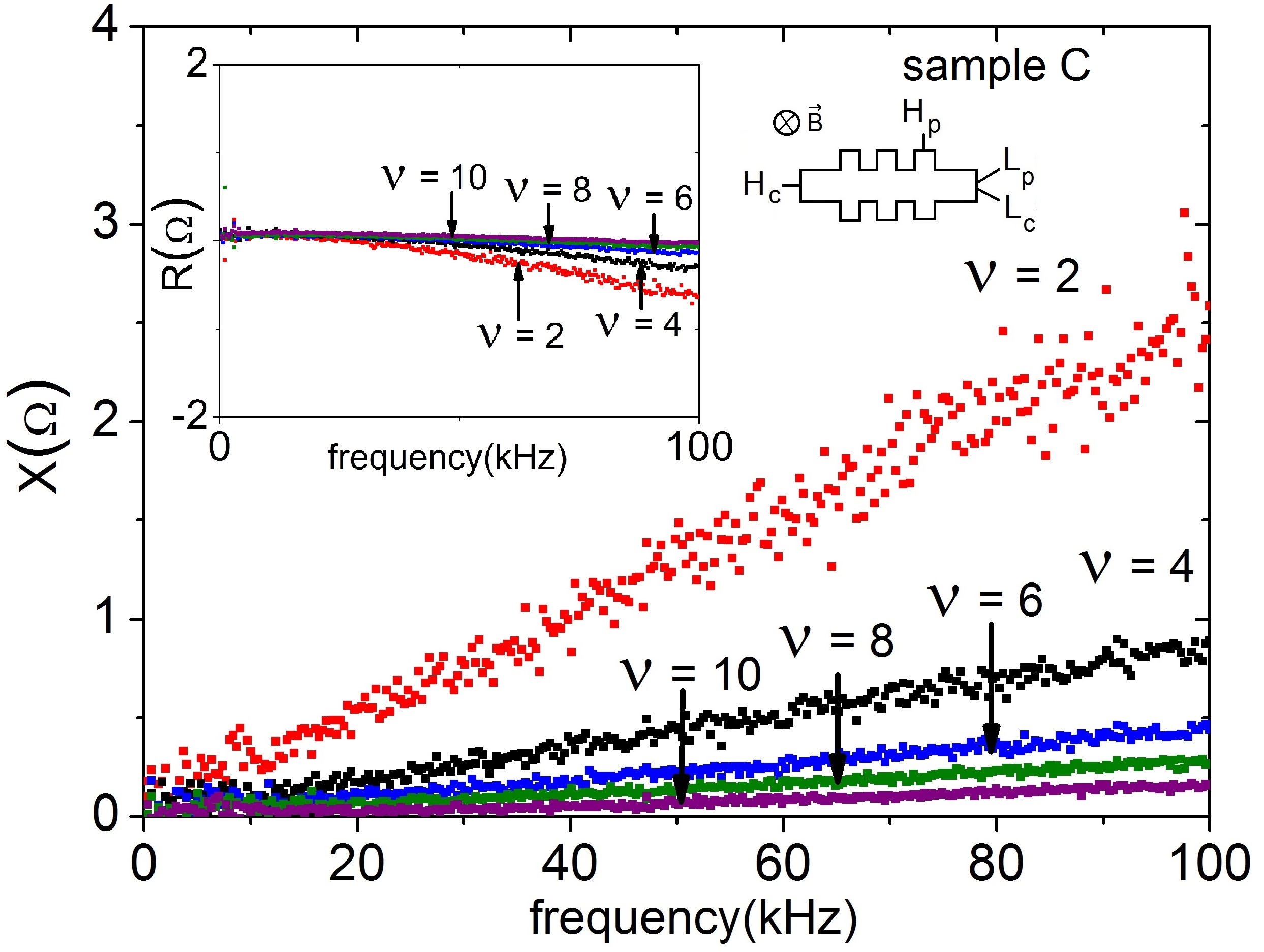}
	\caption{\label{FF2} Reactance as a function of frequency, for
sample C and $\nu=2$ to $10$ by steps of $2$. Configuration is shown in inset.} 
\end{figure}

\begin{figure}
 \centering
	\includegraphics[width=8cm]{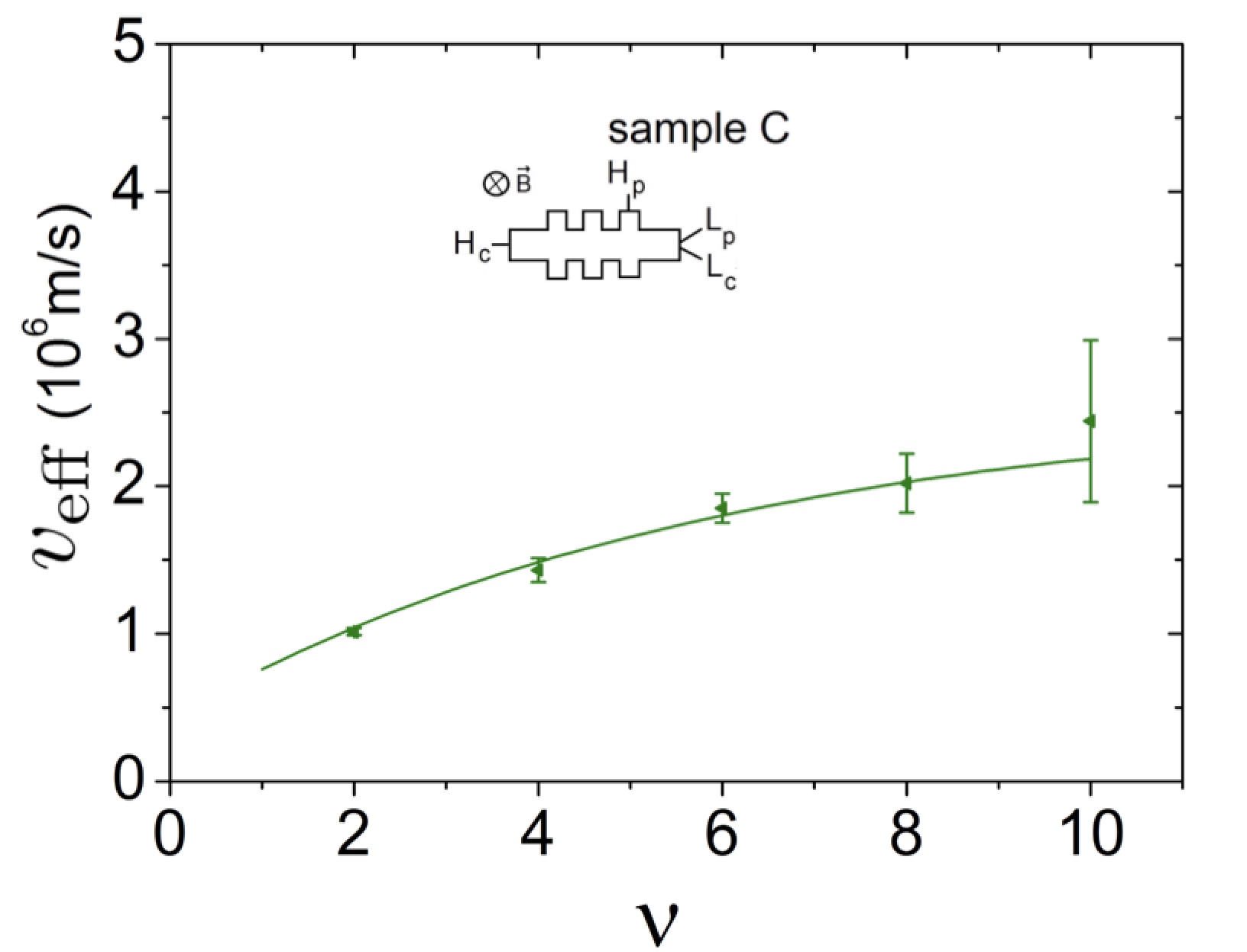}
	\caption{\label{figure-FF2-fit} Plot of the effective velovity
	$v_{\text{eff}}(\nu)$ obtained from the data displayed on
	Fig.~\ref{FF2} and fit by Eq. (1) of the paper using $v_{\text{d}}(\nu)=v_{\text{d}}/\sqrt{\nu}$
	with
	$v_{\text{d}}=\SI{2.2e5}{\meter\second^{-1}}$.
	}
\end{figure}

For Figs. \ref{FF1}, \ref{FF3} and \ref{FF4}, the data have only been
obtained for two values of $\nu$. However, Fig. \ref{FF2} displays all
even values of $\nu$ from $2$ to $10$ for sample C in a given
confuguration. It is then possible to try fitting these experimental
data using Eq.~(1) of the paper with
$v_{\text{d}}(\nu)=v_{\text{d}}/\sqrt{\nu}$. This fit is presented on
Fig. \ref{figure-FF2-fit} and show that our model also described
correctly the experimental data for sample C with the configuration
displayed on Fig.~\ref{FF2}.

%\pagebreak
\subsection{Influence of configurations for other samples}

\begin{figure}
    \centering
    \includegraphics[width=8cm]{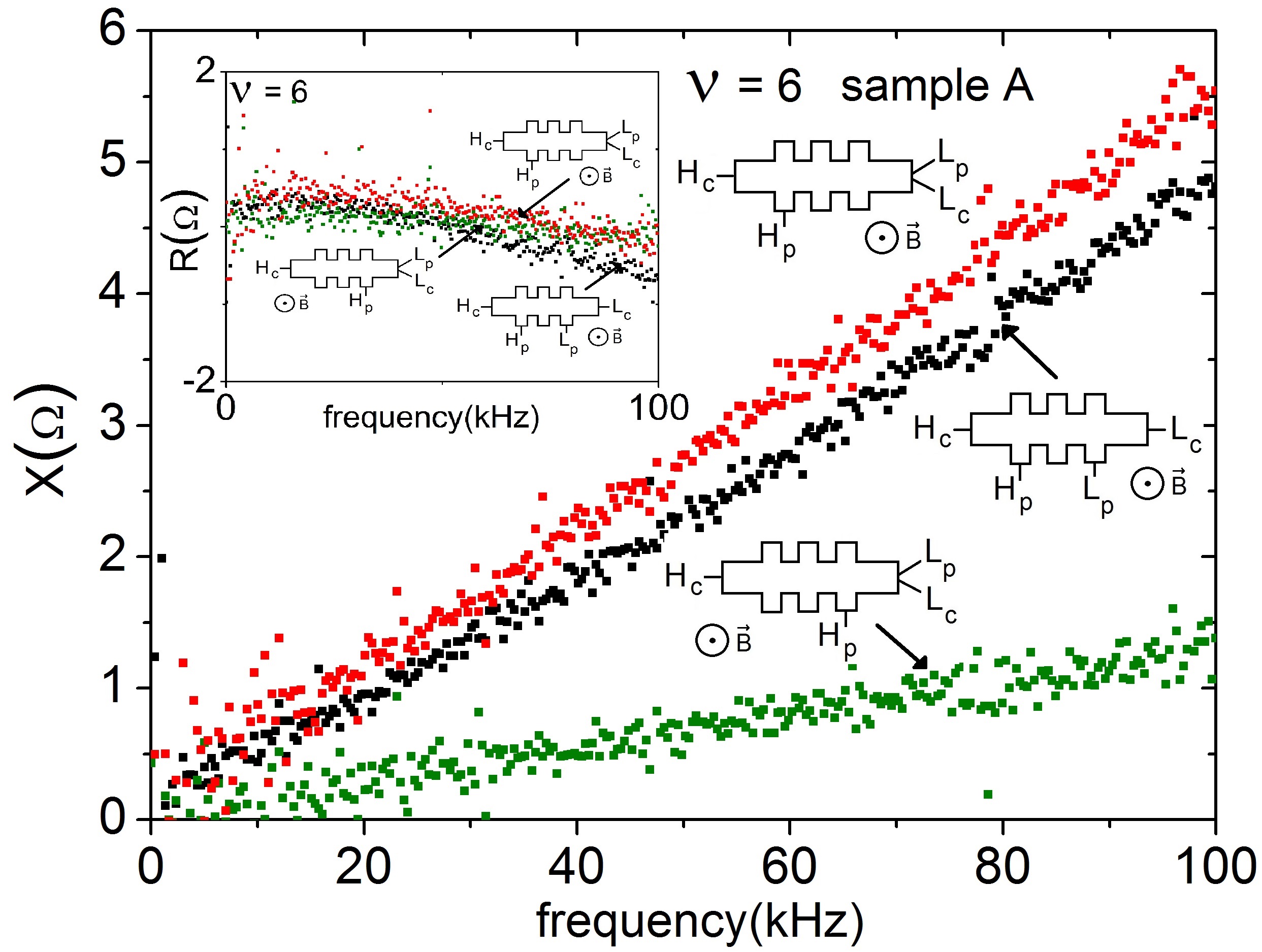}
	\caption{\label{C1} Reactance as a function of frequency, for sample A at $\nu=6$ and for the three configurations.}
\end{figure}

Figs.~\ref{C1} %, \ref{C2}, \ref{C3}, 
to \ref{C4} depict the reactance as a function of frequency for three samples at 
different filling factors.
For each sample at a given filling factor, 
we observe how the configuration (shown as an inset for each curve) modifies the slope of the reactance. 
Each configuration has its proper length and the larger the edge states
the higher the quantum inertia. The length concerned here
is the distance between potentials $L_p$ and $H_p$.  

\begin{figure}
    \centering
    \includegraphics[width=8cm]{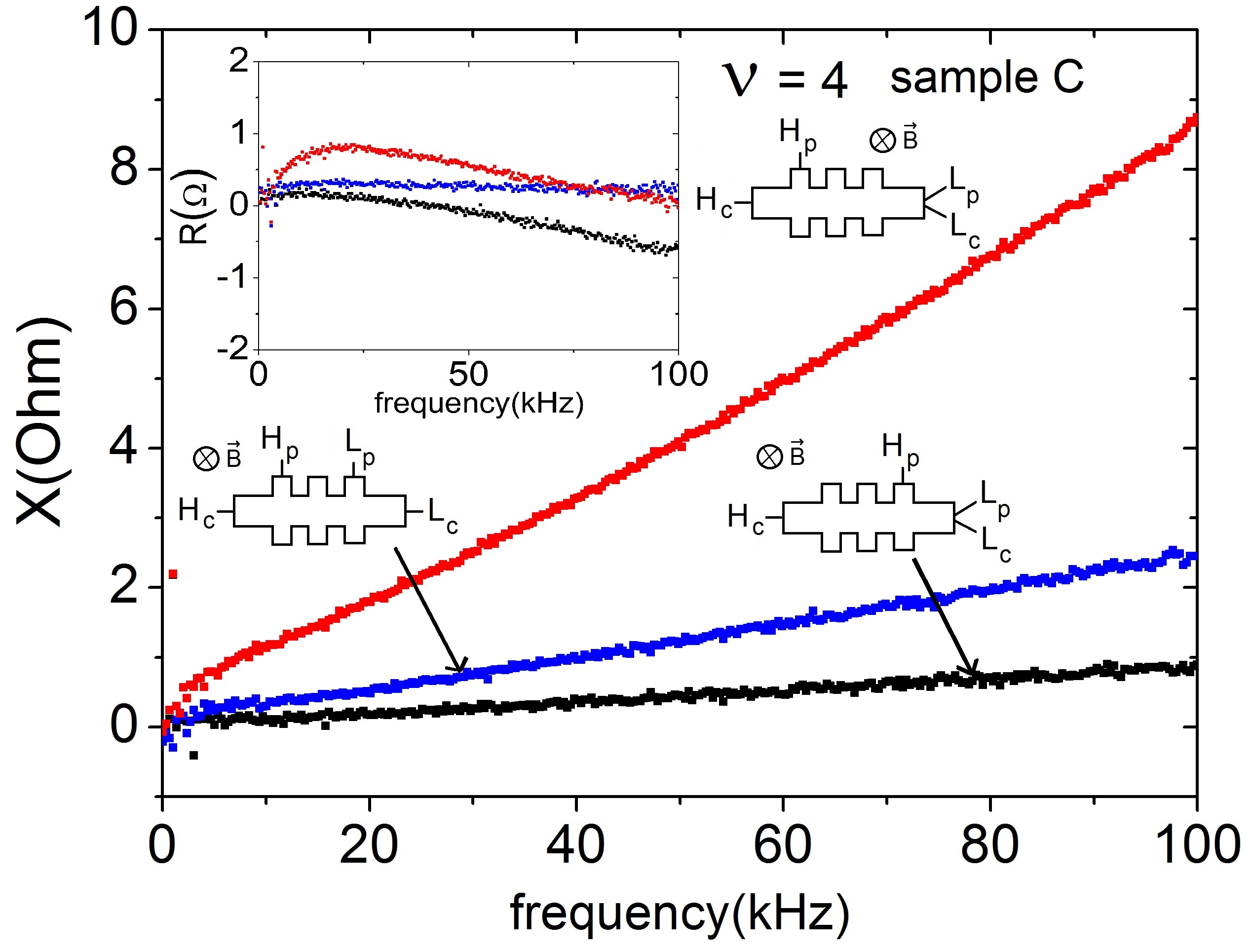}
	\caption{\label{C2} Reactance as a function of frequency, for sample C at $\nu=4$ and for the three configurations.}
\end{figure}
\begin{figure}
    \centering
    \includegraphics[width=8cm]{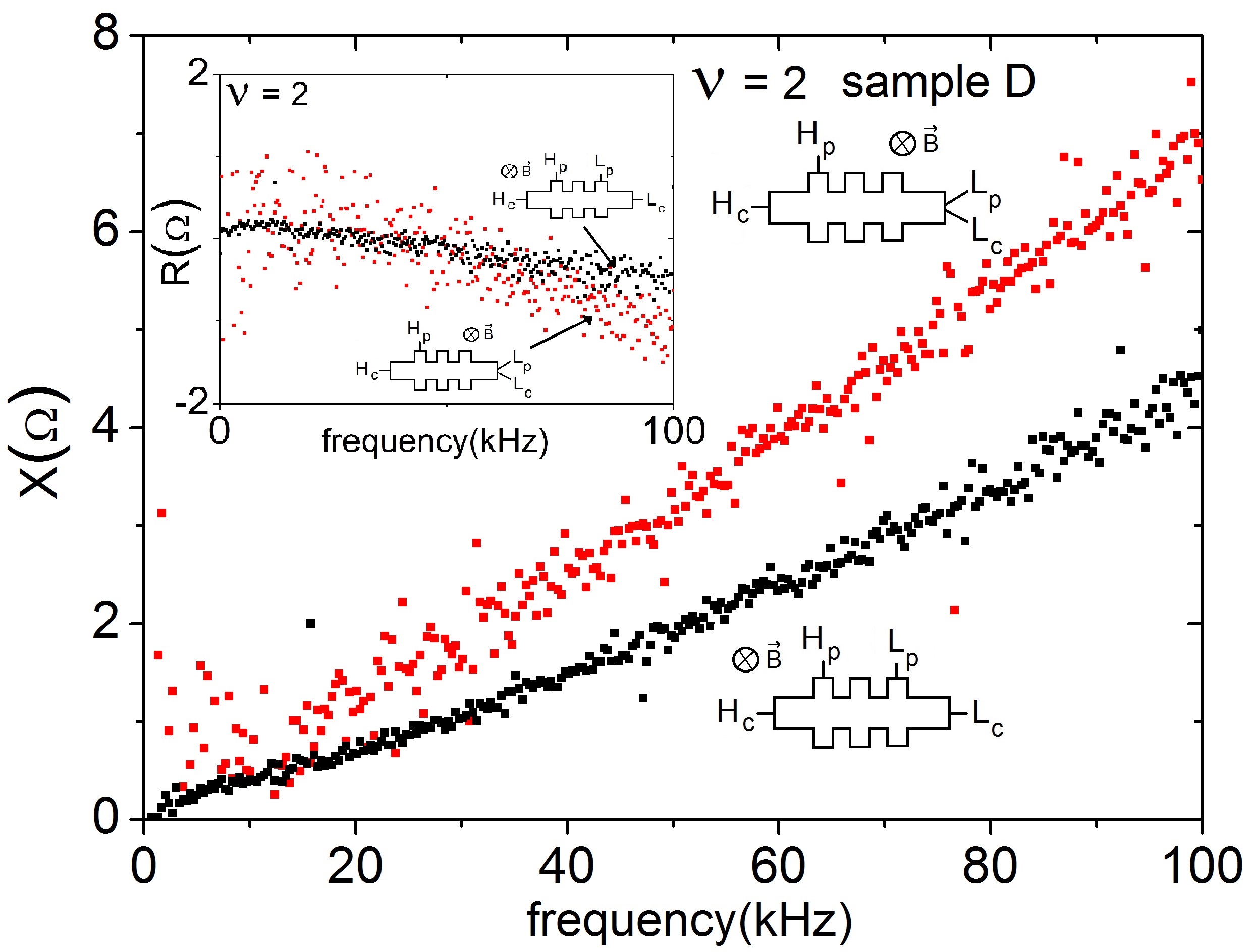}
	\caption{\label{C3} Reactance as a function of frequency for sample D at $\nu=2$ for two configurations.}
\end{figure}

\begin{figure}
    \centering
    \includegraphics[width=8cm]{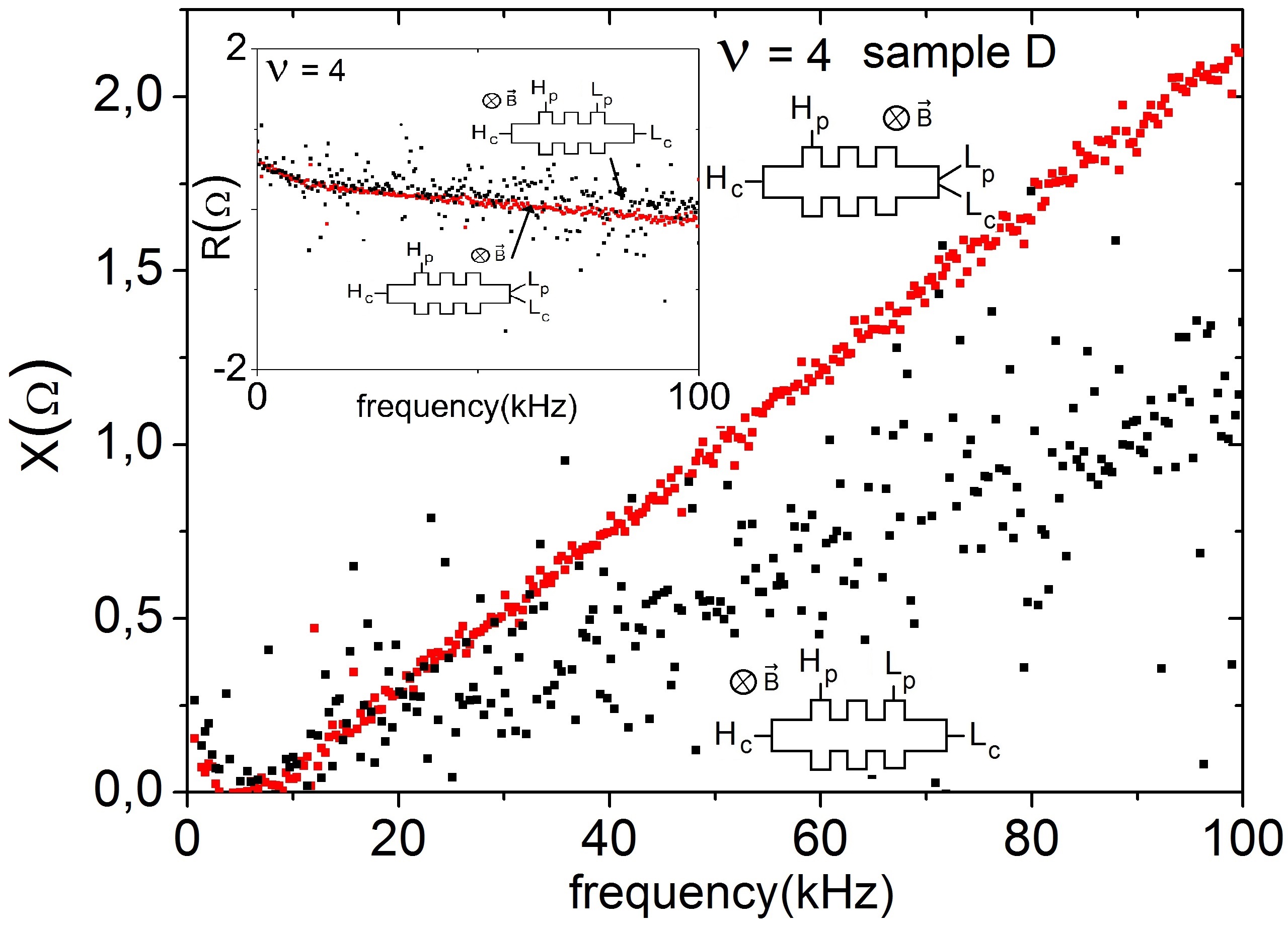}
	\caption{\label{C4} Reactance as a function of frequency for sample
	D at $\nu=4$ for two configurations. Note the smaller values for
	$X(\Omega)$ compared to the other curves which may explain the more
	noisy dataset for the four-terminal geometry.}
\end{figure}

Meanwhile, differences in velocity $v_{\text{d}}$ for distinct 
configurations imply that the ratio between 
quantum inertia and length of edge states not a constant.
Configurations are obtained after  
wire-bonding the samples. This entails heating the samples and cool them down 
again afterwards, and this causes the velocity $v_{\text{eff}}$ to change a bit 
from one configuration to another.

\bibliography{biblio/bigbib,biblio/livres}

\end{document}